\documentclass[twocolumn,notitlepage,prl,superscriptaddress,longbibliography]{revtex4-2}
\usepackage{amsfonts}
\usepackage{textcomp}
\usepackage{times}
\usepackage{graphicx}
\usepackage{float}
\usepackage{latexsym,amsmath,amssymb,bm,euscript}
\usepackage{color}
\usepackage{subfigure}
\usepackage{epstopdf}
\usepackage[colorlinks=true,linkcolor=blue,citecolor=blue]{hyperref}
\usepackage{hyperref}
\usepackage{soul}
\usepackage[normalem]{ulem}
\usepackage{mathrsfs}
\usepackage{amsmath}
\usepackage{xspace}
\usepackage{natbib}
\usepackage{ulem}
\usepackage{etoolbox}
\usepackage{tikz}

\graphicspath{{../Fig/}}

\begin{document}
\title{Dipolar Spin Liquid Ending with Quantum Critical
Point in a Gd-based Triangular Magnet}
	
\author{Junsen Xiang}
\thanks{These authors contributed equally to this work.}
\affiliation{Beijing National Laboratory for Condensed Matter Physics,
Institute of Physics, Chinese Academy of Sciences, Beijing 100190, China}

\author{Cheng Su}
\thanks{These authors contributed equally to this work.}
\affiliation{School of Physics, Beihang University, Beijing 100191, China}

\author{Ning Xi}
\thanks{These authors contributed equally to this work.}
\affiliation{CAS Key Laboratory of Theoretical Physics, Institute of
Theoretical Physics, Chinese Academy of Sciences, Beijing 100190, China}

\author{Zhendong Fu}
\affiliation{Neutron Platform, Songshan Lake Materials Laboratory, Dongguan 523808, China}

\author{Zhuo Chen}
\affiliation{School of Mechanical Engineering, Beijing Institute of Technology, Beijing 100081, China}

\author{Hai Jin}
\affiliation{Department of Astronomy, Tsinghua University, Beijing 100084, China}

\author{Ziyu Chen}
\affiliation{School of Physics, Beihang University, Beijing 100191, China}

\author{Zhao-Jun Mo}
\affiliation{Ganjiang Innovation Academy, Chinese Academy of Sciences,
Ganzhou 341119, People’s Republic of China.}

\author{Yang Qi}
\affiliation{State Key Laboratory of Surface Physics, Fudan University,
Shanghai 200433, China}
\affiliation{Center for Field Theory and Particle Physics, Department
of Physics, Fudan University, Shanghai 200433, China}

\author{Jun Shen}
\affiliation{School of Mechanical Engineering, Beijing Institute of Technology, Beijing 100081, China}
\affiliation{Technical Institute of Physics and Chemistry, Chinese Academy
of Sciences, Beijing 100190, China}

\author{Long Zhang}
\affiliation{Kavli Institute for Theoretical Sciences, and School of Physical
Sciences, University of Chinese Academy of Sciences, Beijng 100049, China}
\affiliation{CAS Center for Excellence in Topological 	Quantum Computation,
University of Chinese Academy of Sciences, Beijng 100190, China}

\author{Wentao Jin}
\email{wtjin@buaa.edu.cn}
\affiliation{School of Physics, Beihang University, Beijing 100191, China}

\author{Wei Li}
\email{w.li@itp.ac.cn}
\affiliation{CAS Key Laboratory of Theoretical Physics, Institute of
Theoretical Physics, Chinese Academy of Sciences, Beijing 100190, China}
\affiliation{CAS Center for Excellence in Topological Quantum Computation,
University of Chinese Academy of Sciences, Beijng 100190, China}
\affiliation{Peng Huanwu Collaborative Center for Research and Education,
Beihang University, Beijing 100191, China}

\author{Peijie Sun}
\email{pjsun@iphy.ac.cn}
\affiliation{Beijing National Laboratory for Condensed Matter Physics,
Institute of Physics, Chinese Academy of Sciences, Beijing 100190, China}

\author{Gang Su}
\email{gsu@ucas.ac.cn}
\affiliation{Kavli Institute for Theoretical Sciences, and
School of Physical Sciences, University of Chinese
Academy of Sciences, Beijng 100049, China}
\affiliation{CAS Center for Excellence in Topological
Quantum Computation, University of Chinese Academy
of Sciences, Beijng 100190, China}

\begin{abstract}
By performing experiment and model studies on a triangular-lattice
dipolar magnet KBaGd(BO$_3$)$_2$ (KBGB), we find the highly
frustrated magnet with a planar anisotropy hosts a strongly fluctuating 
dipolar spin liquid (DSL), which originates from the intriguing interplay 
between dipolar and Heisenberg interactions. The DSL constitutes an 
extended regime in the field-temperature phase diagram, which gets 
lowered in temperature as field increases and eventually ends with an 
unconventional quantum critical point (QCP) at $B_c\simeq 0.75$~T. 
Based on the dipolar Heisenberg model analysis, the DSL is identified 
as a Berezinskii-Kosterlitz-Thouless (BKT) phase with emergent U(1) 
symmetry,  and the end QCP belongs to the 3D XY universality class. 
Due to the tremendous entropy accumulation that can be related to the 
strong BKT and quantum fluctuations, unprecedented magnetic cooling 
effects are observed in the DSL and particularly near the QCP, making 
KBGB a superior dipolar coolant to commercial Gd-based refrigerants. 
We establish the phase diagram for triangular-lattice dipolar quantum 
magnets where emergent symmetry plays an essential role, and provide 
a basis and opens an avenue for their applications in sub-Kelvin refrigeration.
\end{abstract}

\date{\today}
\maketitle

\textit{Introduction.---} Triangular-lattice quantum antiferromagnets
have raised great research interest recently due to the unusual
quantum spin states and transitions therein~\cite{Collins1997,
Starykh2015}. One prominent example is the quantum spin liquid (QSL)
\cite{Anderson1973,Zhou2017,Balents2010} and its possible materialization
in organic compounds~\cite{Shimizu2003,Yamashita2010,Kanoda2011}
and rare-earth triangular magnets~\cite{Li2015YMGO1,Li2015YMGO2,
Shen2016,Paddison2017,Shen2018,Liu2018,Bordelon2019,Dai2021}.
The intriguing spin frustration effects and two dimensionality of such 
systems imply Berezinskii-Kosterlitz-Thouless (BKT) physics may 
appear at low temperatures. Indeed, the Co-based quantum
antiferromagnet Na$_2$BaCo(PO$_4$)$_2$ hosts persistent
spin fluctuations~\cite{Zhong2019,LiN2020,Lee2021,Wellm2021}
till very low temperature, and is proposed to realize spin supersolidity
with prominent BKT phase fluctuations~\cite{Gao2022QMats}. Besides,
emergent symmetry has also been disclosed on the triangular lattice
as a consequence of strong frustration, with a primary example
of rare-earth magnet TmMgGaO$_4$~\cite{Cava2018,Shen2019,
Li2020,Lih2020,ZHu2020,Dun2020neutron}.

% ====== Fig. 1: Material structure & DH model ====== %
\begin{figure}[htp]
\includegraphics[width=1\linewidth]{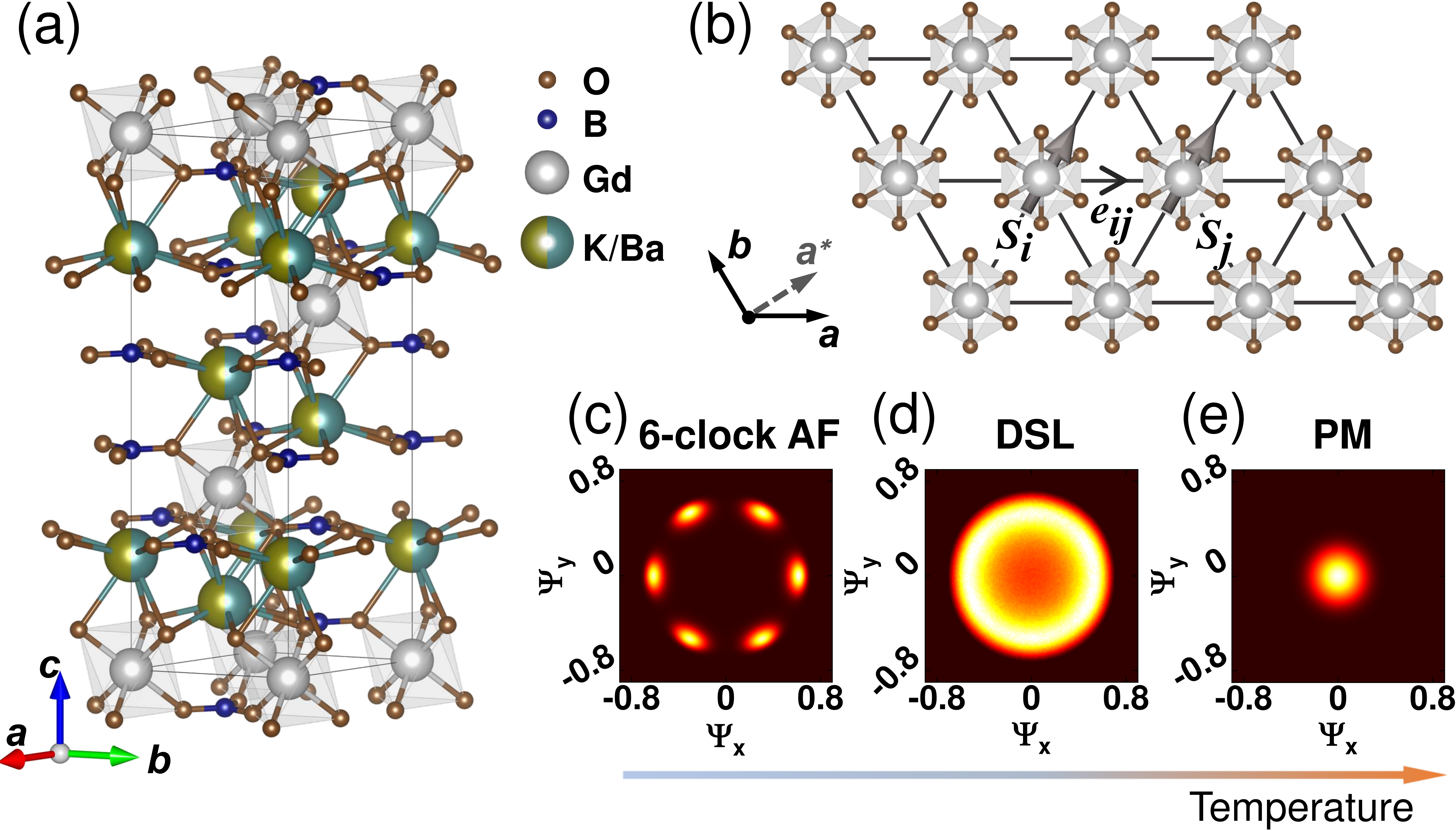}
\caption{(a) shows the crystal structure of KBaGd(BO$_3$)$_2$, and 
(b) the triangular-lattice layers of GdO$_6$ octahedra separated by 
the Ba/K layers with site mixing. The grey arrows refer to the spins 
on site $i$ and $j$, and the unit vector $\bold{e}_{ij}$ is also indicated.
Dipole-dipole interactions are bond-dependent and follow the $\bar 3 m$
site symmetry. (c)-(e) are histograms of the order parameter 
$\Psi_{xy} \equiv \Psi_x + i \Psi_y$ for the 6-clock 
antiferromagnetic (AF) phase, emergent U(1) dipolar 
spin liquid (DSL), and paramagnetic (PM) phase~\cite{SM}.}
\label{Fig1}
\end{figure}
% ================================================ %

Recently, it has been theoretically proposed that the dipolar interactions
can give rise to QSL in triangular-lattice quantum spin systems
\cite{Yao2018}. Lately such dipolar system has been realized
in Yb-based triangular compounds~\cite{Zeng2020,Bag2021,
Cho2021BYBO,Khatua2022Ba3RB9O18,Jiang2022}. However,
the dipolar interactions are rather weak and it is very challenging
for conventional thermodynamic and spectroscopic measurements
to probe the exotic spin states due to dipolar interactions. On the 
contrary, the rare-earth dipolar magnets with even larger moments,
\emph{e.g.}, Gd-based compounds with $\mu_{\rm eff} \approx 8
\mu_B$ and high spin $S=7/2$, are much less explored both in
experiments and theories.
It is expected that the dipolar frustration effects are a priori more
evident in these systems. Moreover, high-spin frustrated systems,
especially those with spin-liquid like behaviors~\cite{Liu2022},
can possess large entropy density and cooling capacity, holding 
thus strong promise as excellent coolants for sub-Kelvin space applications~\cite{Hagmann1995,Shirron2014} and quantum computing~\cite{Jahromi2019nasa}.

In this work, we perform low-temperature thermodynamic and
magnetocaloric measurements on single-crystal samples of
gadolinium borate KBaGd(BO$_3$)$_2$ (KBGB). 
The thermodynamic measurements suggest a dipolar
spin liquid state with no conventional ordering but strong spin
fluctuations as reflected in the algebraic specific heat and
imaginary dynamical susceptibility ($\chi^{''}\rm_{ac}$).
We establish a dipolar Heisenberg model with both dipole-dipole
and Heisenberg interactions for KBGB. Monte Carlo (MC)
simulations well explain the experimental results and unveil
exotic spin states and transitions in the phase diagram.
In particular, the model simulations suggest a two-step melting
of the 6-clock antiferromagnetic (AF) order [c.f., Fig.~\ref{Fig1}(c)]
via two BKT transitions, between which a floating BKT phase 
appears with an emergent U(1) symmetry, well accounting for 
the experimental observations. Consequently, giant magnetocaloric 
effect (MCE) is observed in the quasi-adiabatic demagnetization 
measurements. In particular, we find a clear dip in temperature 
at $B_c \simeq 0.75$~T, i.e., near the quantum critical point 
(QCP) also with emergent U(1) symmetry. 
The obtained lowest temperature of 70~mK clearly surpasses that 
of commercial refrigerant Gd$_3$Ga$_5$O$_{12}$ (GGG) under 
similar conditions, opening an avenue for exploring not only exotic 
spin states and transitions but also superior quantum coolants.

% ====== Fig. 2: Specific heat & magnetic susceptibility ====== %
\begin{figure*}[htp]
\includegraphics[width=1\linewidth]{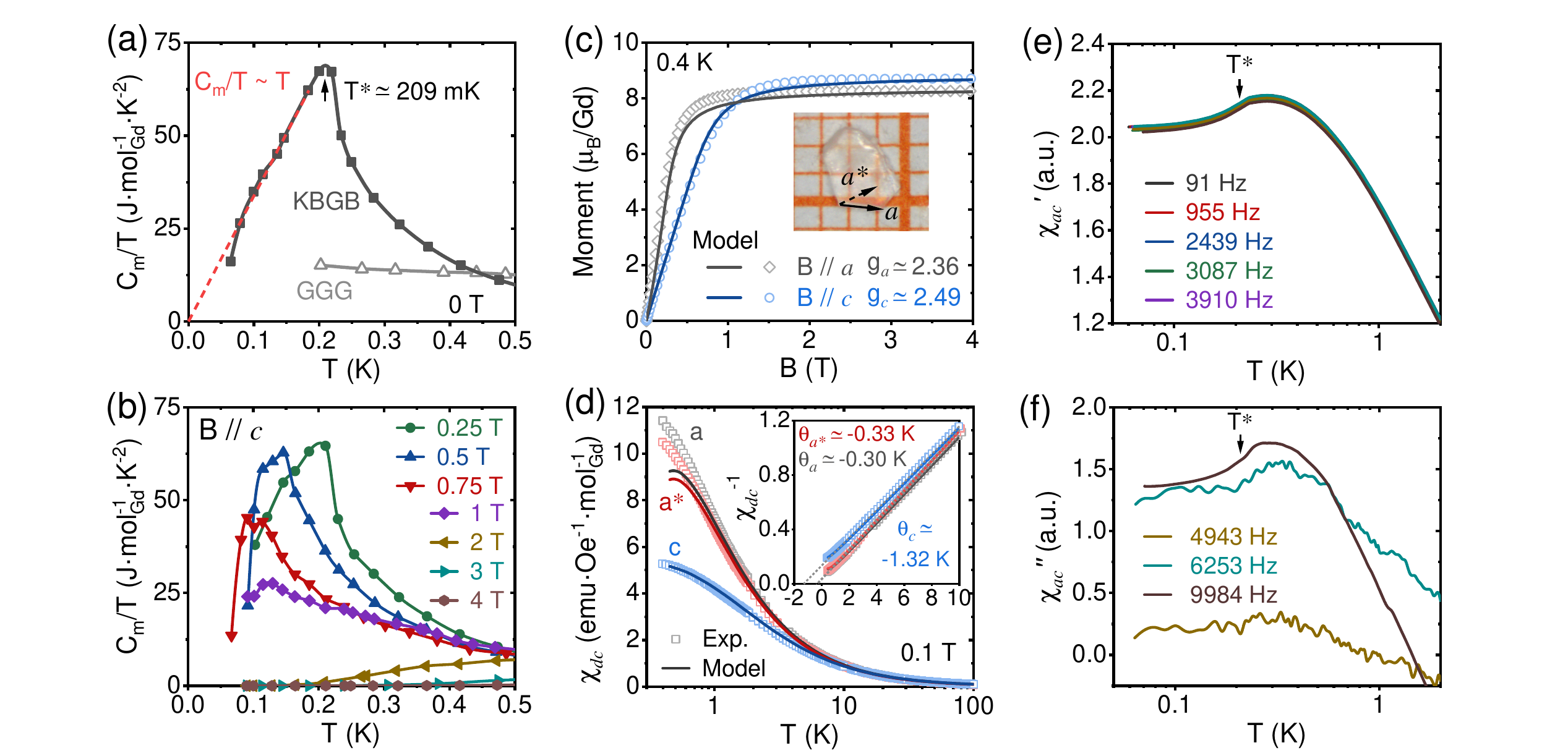}
\caption{Specific heat of KBGB under (a) zero and (b) finite fields
along out-of-plane direction. An algebraic $C_m \sim
T^2$ scaling is observed below the peak temperature $T^*$ in (a),
where the $C_m/T$ values far outweigh that of GGG~\cite{Schiffer1994}.
(c) shows the magnetization curves of the single-crystal KBGB sample
for $B$//$a$ and //$c$, and the results show excellent agreement with
the DH model calculations (solid lines).
The saturation moments are $\mu_a^{\rm sat} \simeq 8.26 \mu_B$ and
$\mu_c^{\rm sat} \simeq 8.72 \mu_B$, from which we determine the
Land\'e factors $g_a\simeq 2.36$ and $g_c\simeq 2.49$, respectively.
The as-grown KBGB single crystal is shown in the inset, with directions
$a$ and $a^*$ also indicated.
(d) shows the molar dc magnetic susceptibilities ($\chi\rm_{dc}$)
measured along the $a$, $a^*$, and $c$ axes, respectively, where
the solid lines representing the DH model calculations show excellent
agreements. The inset shows the Curie-Weiss fittings in the
paramagnetic regime 0.4~K $\leq T \leq$ 10~K, with the fitted
Curie-Weiss temperatures $\theta_{a, a^*, c}$ also indicated.
(e, f) present respectively the real and imaginary ac susceptibilities
measured with different frequencies.
}
\label{Fig2}
\end{figure*}

\textit{Crystal structure and effective model for KBaGd(BO$_3$)$_2$.---}
Centimeter-sized single crystals of KBGB were synthesized using the
flux method as described in detail in Supplementary Materials (SM)
\cite{SM}. X-ray diffraction measurements indicate high quality
of the single crystals, and confirm the trigonal structure
\cite{Sanders2017,Guo2019KBYB} with space group $R$-3$m$
[\emph{c.f.}, Fig.~\ref{Fig1}(a)]. Magnetic Gd$^{3+}$ ions with 4$f^7$
electron configuration ($L= 0, S=7/2$) form perfect triangular lattice
[Fig.~\ref{Fig1}(b)], with a relatively high ionic density of 6.4~nm$^{-3}$.

The direct dipolar interaction between magnetic ions Gd$^{3+}$ has 
a characteristic energy $E_{\rm dp} \sim 2\mu_{0}\mu_{\rm sat}^2 /
4 \pi a^{3} \approx 0.05$~meV (with $\mu_{\rm sat} \approx 8$
$\mu\rm_B$), which determines the low-temperature spin states
in KBGB. To simulate the dipolar magnet, we consider the 
Hamiltonian ${H}=J_{H}\sum_{\langle i,j \rangle_{\rm NN}}
\mathbf{S}_{i}\cdot\mathbf{S}_{j}+J_{D}\sum_{ i,j} [{\mathbf{S}_{i} \cdot
\mathbf{S}_{j}}-{3(\mathbf{S}_{i}\cdot\mathbf{e}_{ij})(\mathbf{S}_{j} \cdot
\mathbf{e}_{ij})}]/r_{ij}^{3}$, where $\mathbf{e}_{ij}$($r_{ij}$) refers to 
the unit vector(distance) between site $i$ and $j$ in the unit of lattice 
constant $a$. $J_{H}$ and $J_{D}$ refer to the nearest neighbor (NN)
Heisenberg and dipole-dipole interactions, respectively. As the dipolar
interactions show rapid (cubic) power-law decay with longer range
interactions washed out, below we keep only NN terms
\begin{equation}
{H}_{\rm DH}= \sum_{\langle i,j\rangle_{\rm NN}} J \, \mathbf{S}_{i}
\cdot\mathbf{S}_{j}-D\,(\mathbf{S}_{i}\cdot\mathbf{\it e}_{ij})
(\mathbf{S}_{j}\cdot\mathbf{\it e}_{ij}),
\label{Eq:DHM}
\end{equation}
where $J=J_{H}+J_{D}$ is the NN isotropic coupling and $D=3 J_{D}$
refers to the dipolar anisotropic term. We perform MC simulations of 
the NN dipolar Heisenberg (DH) model on up to $60 \times 60$ 
triangular lattice~\cite{SM}, and find the results fit very well the 
experimental results. The determined coupling parameters from 
the fittings are $J\simeq47$~mK and $D\simeq80$~mK, which 
correspond to $J_D \simeq 27$~mK and $J_H\simeq 20$~mK, 
leading to a dipolar energy of about 660~mK. Such estimate based 
on MC fittings is consistent with the direct interaction of $E_{\rm dp} 
\approx 0.05$~meV evaluated above.

\textit{Magnetic specific heat, susceptibility, and dipolar spin liquid.---} 
In Fig.~\ref{Fig2}(a) we show the zero-field specific heat $C_m$ 
measured down to 65~mK for KBGB. There exists a round peak 
at $T^*\simeq 209$~mK, below which the system exhibits $C_m 
\sim T^2$ with algebraic scaling, resembling that of two-dimensional 
(2D) Heisenberg or XY quantum spin model with U(1) symmetry
\cite{Cv_Hasenfratz_1993,Cv_Sandvik1999}. The thermodynamic
measurements suggest a gapless liquid-like and strongly fluctuating
spin state as if the dipolar planar anisotropy were absent. 
This is also 
reflected in the huge low-temperature specific heat in KBGB,
which far exceeds that of the renowned Gd-based refrigerant GGG
\cite{Schiffer1994,paddison2015hidden,Shirron2014}.

\begin{figure}[htp]
\includegraphics[width=1\linewidth]{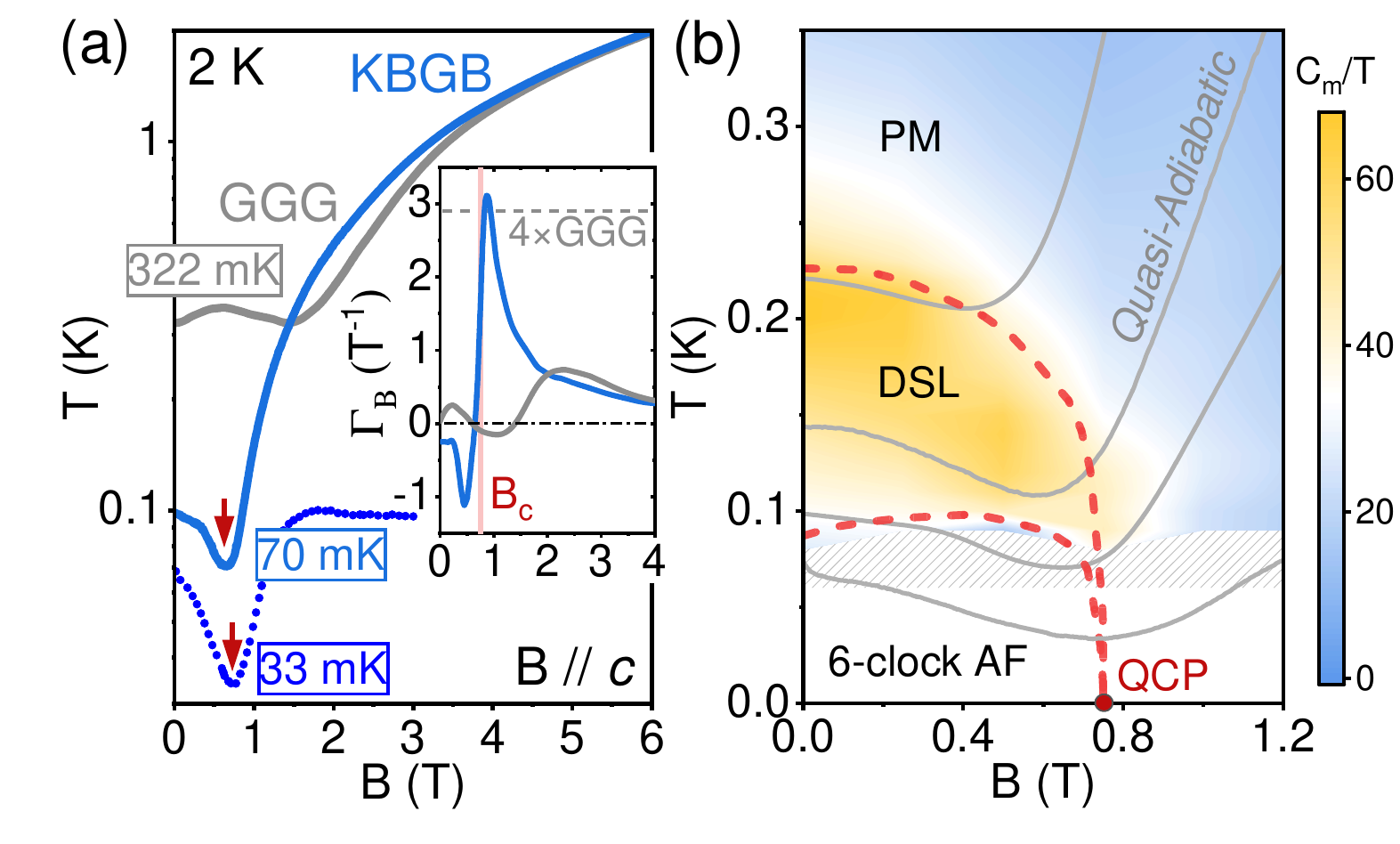}
\caption{(a) shows the quasi-adiabatic isentropes measured in KBGB
under out-of-plane fields. The KBGB curve
exhibits a clear dip at the lowest temperature $T_m \simeq 70$~mK,
much lower than that of GGG ($T_m\simeq 322$~mK). Starting from
$T_i\simeq95$~mK, KBGB can reach a remarkably low temperature
$T_m\simeq 33$~mK in the dip (blue dotted line). The inset shows the
magnetic Gr\"uneisen ratio $\Gamma_B$ deduced from the curves in 
the main panel.
(b) shows the phase diagram of KBGB with the $C_m/T$ contour plot in
the background. The bright regime with large spin fluctuations represent 
the DSL, with schematic dashed line boundaries, ending with a QCP at
$B_c \simeq 0.75$~T.}
\label{Fig3}
\end{figure}

In Fig.~\ref{Fig2}(b), we apply out-of-plane fields ($B$//$c$) to the
compound, and find the round $C_m$ peaks move towards lower 
temperature with heights slightly reduced. This suggests that the 
spin liquid states constitute an extended phase that we dub as 
\textit{dipolar spin liquid} (DSL).
As field further increases and exceeds about 0.75~T, the DSL behavior 
disappears [\emph{c.f.}, the contour plot of $C_m$/$T$ in Fig.~\ref{Fig3}(b)], 
and the $C_m$ peaks move instead to the high-temperature side with 
low-energy fluctuations quickly suppressed.

Fig.~\ref{Fig2}(c) shows the isothermal magnetization measured at
$T$ = 0.4 K, where a clear magnetic anisotropy between the out-of-plane 
(//$c$ axis) and in-plane (//$a$) directions is observed.
This anisotropy can be clearly recognized in the different saturation
magnetization moments, and the transition fields are also different
along two directions, \textit{i.e.}, about 1~T(0.5~T) along $c$($a$) 
axis. Likewise, the low-temperature dc susceptibility ($\chi_{\rm dc}$)
exhibits a clear easy-plane anisotropy, see Fig.~\ref{Fig2}(d). 
Although the determined Land\'e factor $g_c \simeq 2.49$ is slightly 
larger than $g_a\simeq 2.36$, the intrinsic dipolar anisotropy leads to 
larger in-plane $\chi_{\rm dc}$ (along $a$ and $a^*$ axes) than that 
along the $c$ axis. The negative Curie-Weiss temperatures fitted from 
the dc susceptibility reflect the AF nature, and the slightly different 
$\theta_a\simeq -300$~mK and $\theta_{a^*}\simeq -330$~mK 
reveal the planar anisotropy. This small but sensible planar 
anisotropy between $a$ and $a^*$ axes can be ascribed to the 
bond-dependent dipolar interaction [\emph{c.f.}, Eq.~(\ref{Eq:DHM})].

The proposed DSL is further corroborated by the ac magnetic 
susceptibilities shown in Figs.~\ref{Fig2}(e,f). The real part 
$\chi'_{\rm ac}$ exhibits a frequency-independent maximum 
and remains large even below the characteristic temperature 
$T^*$. The spin-glass scenario is therefore firmly excluded 
despite the K/Ba site mixing in the compound. Interestingly, 
the imaginary ac susceptibility $\chi''_{\rm ac}(T)$, although 
being featureless for low frequencies $\omega \lesssim 4$~kHz,
show a clear temperature-dependent behavior for higher 
frequencies in Fig.~\ref{Fig2}(f). Considering that $\chi''(\omega)$ 
can be directly related to the dynamical correlation $S(\omega)$ 
through the fluctuation-dissipation theorem, $\chi''(\omega)
\propto \frac{\omega}{T} S(\omega)$ ($\omega\ll T$), this 
clearly suggests the persistence of low-energy spin fluctuations 
even below $T^*$ and supports the spin-liquid scenario.
	
\textit{Magnetocaloric effect and quantum critical point.---}
In Fig.~\ref{Fig3}(a), we show the quasi-adiabatic demagnetization 
measurements (see details in SM~\cite{SM}), 
where the lowest temperature $T_m$ is achieved 
around the dip $B_c \simeq 0.75$~T in the isentropic line and 
remains at very low temperature for $B < B_c$ in the small field 
side. In particular, it is found that KBGB clearly outperforms GGG 
in the obtained lowest temperature, \emph{i.e.}, $T_{m} \simeq 
70$~mK (KBGB) vs. 322~mK (GGG), for the same initial condition 
of $T_i = 2$~K and $B_i=6$~T. In Fig.~\ref{Fig3}(b) we provide 
more of the isentropic lines from different initial conditions, and 
observe the highly asymmetric isentropes, which ``levels off'' in 
the bright DSL regime with strong fluctuations reflected in  
large $C_m/T$. 

Remarkably, strong cooling effects are also observed for very 
low, sub-Kelvin initial temperature. In Fig.~\ref{Fig3}(b), we 
obtain a lowest $T_m \simeq 33$~mK from initial $T_i \simeq 95$
mK. Such unprecedented MCE response strongly corroborates 
the existence of QCP at $B_c\simeq0.75$~T, as is further 
substantiated by an evident peak-dip structure with sign change 
in the magnetic Gr\"uneisen ratio $\Gamma_B=\frac{1}{T}
(\frac{\partial T}{\partial B})_S$~\cite{Zhu2003,Xiang2017,Garst2005,
Liu2021}, shown in Fig.~\ref{Fig3} inset, which has been widely 
used in characterizing QCP for heavy fermions~\cite{Tokiwa2009,
Jang2015,Tokiwa2016,Gegenwart2016,shimura2022magnetic} 
and low-dimensional quantum spin systems
\cite{Honecker2009,Wolf2011,Lang2012,Bachus2020}.
The peak height of $\Gamma_B$ exceeds 4 times that 
of GGG, indicating a giant QCP cooling effect in KBGB.

\textit{Emergent symmetry in KBGB.---}
According to the above measurements, we obtain the phase 
diagram of KBGB in Fig.~\ref{Fig3}(b). 
The two schematic dashed lines, enclosing the DSL with large 
$C_m/T$ values, meet at a QCP where the lowest cooling temperature 
is reached. Besides QCP, within the DSL regime we find persistent 
spin fluctuations and cooling effects whose origin is clarified through 
a model analysis below.

% ==== Fig. 3: Low-T MCE measurements and QCPs ==== %
% ============================================ %

%As the model is highly frustrated in the out-of-plane direction,
%the order parameter lies within the $ab$ plane. We conduct MC calculations of the DH model [Eq.~(\ref{Eq:DHM})] for KBGB.

In Figs.~\ref{Fig2}(c,d), we find the
anisotropic susceptibility and magnetization measured along $a$ and
$c$ axes can be well captured by the DH model. The slight deviation
between model calculations and experimental in-plane susceptibilities 
at low temperature $\lesssim0.5$~K may be attributed to strong 
quantum fluctuations of planar order parameters. Besides, 
the model calculations find a specific heat peak at about 270~mK, 
which also gets suppressed as field increases, well resembling the 
experimental results in Figs.~\ref{Fig2}(a,b). As various phases with 
prominent features are well captured by MC simulations~\cite{SM},
we thus confirm that the DH model can very well describe the dipolar 
magnet KBGB.

To characterize the spin states in the phase diagram, we introduce
the order parameter $\Psi_{xy} \equiv m e^{i\theta} = \sum_{j}
e^{i Q r_j} (m_j^x + i m_j^y)$, where $j$ runs over the lattice sites
and $Q = \pm \frac{1}{2} a^*, \pm \frac{1}{2} b^*, \pm \frac{1}{2}
(a^* - b^*)$~\cite{SM}. Histogram of the complex order parameter
$\Psi_{xy}$ at various temperature are shown in Figs.~\ref{Fig1}(c-e).
At low temperature, the dipolar system exhibits a 6-clock AF order
corresponding to $\theta = 0, \pm \pi/3, \pm 2\pi/3$, and $\pi$
\cite{SM}. As temperature ramps up, the six points in the histogram
prolong and merge into a circle with emergent U(1) symmetry,
where the angle $\theta$ can choose arbitrary angle. As temperature
further enhances, the amplitude $m$ eventually vanishes and the
system enters the conventional PM phase. 

Recall that the 6-state
clock model with $\cos{(6\theta)}$ anisotropic term undergoes two 
successive BKT transitions~\cite{Jose1977}, between which the 
anisotropic term becomes irrelevant perturbation, the intermediate
DSL thus constitutes a BKT phase with emergent U(1) symmetry \cite{Moessner2000,Moessner2001,Isakov2003,KTDP_Baek2011PRB}.
For the zero-temperature QCP, as the clock term is dangerously 
irrelevant there~\cite{Moessner2000}, transition occurs directly 
between the 6-clock AF and PM phases and belongs to the 3D 
XY universality class. The corresponding critical exponents
$z=1$ and  $\nu \simeq 0.66$~\cite{Gottlob1993} can lead to
a strongly diverging $\Gamma_B\sim T^{-1/z \nu}$~\cite{Zhu2003}, 
which may account for the very sharp peak observed in 
Fig.~\ref{Fig3}(a) inset. Therefore, emergent symmetry 
constitutes a key for understanding the spin liquid and 
quantum criticality in KBGB.

%. ======= Fig. 4 ========%
\begin{figure}[thp]
\includegraphics[width=1\linewidth]{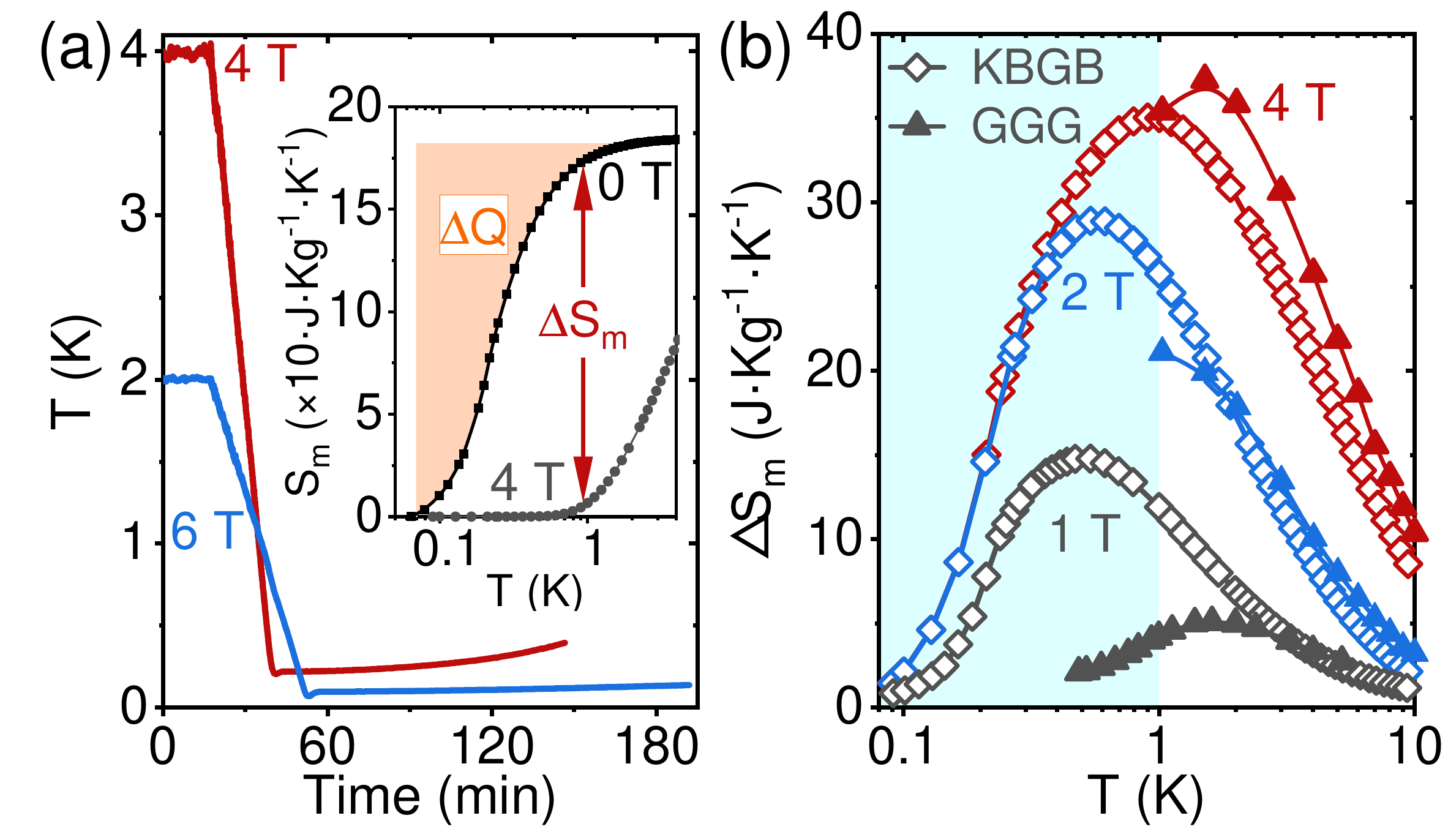}
\caption{(a) The quasi-adiabatic demagnetization cooling curves of
KBGB single-crystal sample (0.5~g), starting from two different initial 
conditions ($T_i=4$~K, $B_i=4$~T) and ($T_i=2$~K, $B_i=6$~T), 
and reaching lowest temperature $T_m\simeq 205$~mK and $70$~mK, 
respectively.
Parasitic heat loads are estimated to be 0.2~$\mu$W for $T_i=4$~K 
environment and 0.05~$\mu$W for $T_i=2$~K. The inset shows 
magnetic entropy under zero and 4~T fields, with the shaded area 
representing the absorbed heat $\Delta Q= 47.44~$J$\cdot$Kg$^{-1}$ 
in the hold process. (b) plots the entropy change results $\Delta S_m$ 
vs. $T$, for fields decreasing from various peak values to zero, and
compared to those of GGG~\cite{Schiffer1994,GLF2006}.
}
\label{Fig4}
\end{figure}

% =========== Performance ========== %
\textit{Superior cooling performance.---}
Starting from $T_i$ = $2$~K, KBGB single crystals are observed
to reach $T_m\simeq70$~mK [Fig.~\ref{Fig3}(a)], far surpassing 
other Gd-based refrigerants, \emph{e.g.}, GGG (322~mK) and 
GdLiF$_4$ (480~mK)~\cite{ADRdesign2014}. Besides, powder 
samples can also achieve much lower $T_m$ than that of GGG
\cite{SM}. Long hold time $t_h$ is also witnessed in KBGB. In the 
environment temperature of 2~K, it remains below 140~mK 
for $t_h\approx 2$~h after the field is exhausted, which can be 
ascribed to the large heat absorption $\Delta Q$ shown in inset 
of Fig.~\ref{Fig4}(a).
The isothermal entropy change $\Delta S_m$ characterizes the cooling
capacity of refrigerants. In Fig.~\ref{Fig4}(b), we compare $\Delta S_m$
of KBGB with that of GGG, and find that KBGB has significantly larger 
$\Delta S_m$ below 1~K [shaded regime in Fig.~\ref{Fig4}(b)], i.e., 
the temperature window of central interest for sub-Kelvin application.
Overall, the low cooling temperature $T_m$, long hold time $t_h$, 
and giant entropy change $\Delta S_m$ suggest that KBGB is a 
superior quantum magnet coolant for cryogenics.

\textit{Discussions and outlook.---}
The pursue for high entropy density and low ordering temperature
constitutes two opposing factors hard to fulfill simultaneously for 
sub-Kelvin refrigerants. Here we find the spin frustration and quantum 
criticality in the dipolar system come to the rescue. We show that 
the compound KBaGd(BO$_3$)$_2$ with high density Gd$^{3+}$
crystallizing on a trianglar lattice is demonstrated to host a disordered 
and strongly fluctuating spin liquid till very low temperature, offering 
enormous cooling capacity. This can be ascribed to the prominent 
BKT fluctuations and quantum criticality in this $S=7/2$ dipolar 
magnet. Despite a planar anisotropy, U(1) symmetry nevertheless
emerges in the compound as revealed by the DH model analysis. 
Although in the present study only NN terms are considered, 
inclusion of further neighboring couplings is believed not to 
change the conclusion here, as it maintains the universality class of 
BKT transitions in the planar dipolar models~\cite{KTDP_Baek2011PRB,
KTDP_Vasiliev2014NJP}.

The scenario of DSL ending up with emergent U(1) 
QCP may also be applicable to other dipolar quantum magnets. 
Recent progress in experimental studies reveal a series of rare-earth
triangular quantum dipolar antiferromagnets, \emph{e.g.},
Ba$_3$REB$_3$O$_9$/Ba$_3$REB$_9$O$_{18}$ (with RE
a rare-earth ion)~\cite{Cho2021BYBO,Khatua2022Ba3RB9O18}
and ABaRE(BO$_3$)$_2$ (with A an alkali ion)~\cite{Guo2019NBYB,
Tokiwa2021}. For example, it has been observed that in
Ba$_3$YbB$_3$O$_9$ that 80\% entropy remain below 56~mK
\cite{Bag2021}, despite a dipolar interaction of about 160~mK, 
suggesting that the DSL and unconventional QCP may also be 
relevant in the Yb-based dipolar compounds. This work, therefore, 
opens a venue for hunting exotic spin states as well as superior 
quantum coolants in triangular dipolar magnets.

\textit{Note added.---} Upon finishing the present work, we are
aware of a recent work~\cite{Jesche2022arXiv} also that also 
conducts the MCE study of KBGB with however polycrystalline 
samples, where they find strong cooling effect down to 121~mK.

\begin{acknowledgments}
\textit{Acknowledgements.---}
W.L. is indebted to Yuan Wan and Tao Shi for helpful discussions.
W.J. and C.S. acknowledge the support from the beamline 1W1A
of the Beijing Synchrotron Radiation Facility. This work was
supported by the National Natural Science Foundation of China
(Grant Nos.~12222412, 11834014, 11974036, 12047503, 12074023,
12074024, 12174387, and 12141002), National Key R \& D Program
of China (Grant No.~2018YFA0305800), Strategic Priority Research
Program of CAS (Grant No.~XDB28000000), and CAS Project for
Young Scientists in Basic Research (Grant No.~YSBR-057). We thank
the HPC-ITP for the technical support and generous allocation of
CPU time. This work was supported by the Synergetic Extreme
Condition User Facility (SECUF).
\end{acknowledgments}

%  ====== Bib ======= %
\bibliography{MCERef.bib}

\begin{thebibliography}{72}%
\makeatletter
\providecommand \@ifxundefined [1]{%
 \@ifx{#1\undefined}
}%
\providecommand \@ifnum [1]{%
 \ifnum #1\expandafter \@firstoftwo
 \else \expandafter \@secondoftwo
 \fi
}%
\providecommand \@ifx [1]{%
 \ifx #1\expandafter \@firstoftwo
 \else \expandafter \@secondoftwo
 \fi
}%
\providecommand \natexlab [1]{#1}%
\providecommand \enquote  [1]{``#1''}%
\providecommand \bibnamefont  [1]{#1}%
\providecommand \bibfnamefont [1]{#1}%
\providecommand \citenamefont [1]{#1}%
\providecommand \href@noop [0]{\@secondoftwo}%
\providecommand \href [0]{\begingroup \@sanitize@url \@href}%
\providecommand \@href[1]{\@@startlink{#1}\@@href}%
\providecommand \@@href[1]{\endgroup#1\@@endlink}%
\providecommand \@sanitize@url [0]{\catcode `\\12\catcode `\$12\catcode
  `\&12\catcode `\#12\catcode `\^12\catcode `\_12\catcode `\%12\relax}%
\providecommand \@@startlink[1]{}%
\providecommand \@@endlink[0]{}%
\providecommand \url  [0]{\begingroup\@sanitize@url \@url }%
\providecommand \@url [1]{\endgroup\@href {#1}{\urlprefix }}%
\providecommand \urlprefix  [0]{URL }%
\providecommand \Eprint [0]{\href }%
\providecommand \doibase [0]{https://doi.org/}%
\providecommand \selectlanguage [0]{\@gobble}%
\providecommand \bibinfo  [0]{\@secondoftwo}%
\providecommand \bibfield  [0]{\@secondoftwo}%
\providecommand \translation [1]{[#1]}%
\providecommand \BibitemOpen [0]{}%
\providecommand \bibitemStop [0]{}%
\providecommand \bibitemNoStop [0]{.\EOS\space}%
\providecommand \EOS [0]{\spacefactor3000\relax}%
\providecommand \BibitemShut  [1]{\csname bibitem#1\endcsname}%
\let\auto@bib@innerbib\@empty
%</preamble>
\bibitem [{\citenamefont {Collins}\ and\ \citenamefont
  {Petrenko}(1997)}]{Collins1997}%
  \BibitemOpen
  \bibfield  {author} {\bibinfo {author} {\bibfnamefont {M.~F.}\ \bibnamefont
  {Collins}}\ and\ \bibinfo {author} {\bibfnamefont {O.~A.}\ \bibnamefont
  {Petrenko}},\ }\bibfield  {title} {\bibinfo {title} {Review/synthèse:
  Triangular antiferromagnets},\ }\href {https://doi.org/10.1139/p97-007}
  {\bibfield  {journal} {\bibinfo  {journal} {Can. J. Phys.}\ }\textbf
  {\bibinfo {volume} {75}},\ \bibinfo {pages} {605} (\bibinfo {year}
  {1997})}\BibitemShut {NoStop}%
\bibitem [{\citenamefont {Starykh}(2015)}]{Starykh2015}%
  \BibitemOpen
  \bibfield  {author} {\bibinfo {author} {\bibfnamefont {O.~A.}\ \bibnamefont
  {Starykh}},\ }\bibfield  {title} {\bibinfo {title} {Unusual ordered phases of
  highly frustrated magnets: a review},\ }\href
  {http://stacks.iop.org/0034-4885/78/i=5/a=052502} {\bibfield  {journal}
  {\bibinfo  {journal} {Rep. Prog. Phys.}\ }\textbf {\bibinfo {volume} {78}},\
  \bibinfo {pages} {052502} (\bibinfo {year} {2015})}\BibitemShut {NoStop}%
\bibitem [{\citenamefont {Anderson}(1973)}]{Anderson1973}%
  \BibitemOpen
  \bibfield  {author} {\bibinfo {author} {\bibfnamefont {P.~W.}\ \bibnamefont
  {Anderson}},\ }\bibfield  {title} {\bibinfo {title} {Resonating valence
  bonds: A new kind of insulator?},\ }\href
  {https://doi.org/https://doi.org/10.1016/0025-5408(73)90167-0} {\bibfield
  {journal} {\bibinfo  {journal} {Mater. Res. Bull.}\ }\textbf {\bibinfo
  {volume} {8}},\ \bibinfo {pages} {153 } (\bibinfo {year} {1973})}\BibitemShut
  {NoStop}%
\bibitem [{\citenamefont {Zhou}\ \emph {et~al.}(2017)\citenamefont {Zhou},
  \citenamefont {Kanoda},\ and\ \citenamefont {Ng}}]{Zhou2017}%
  \BibitemOpen
  \bibfield  {author} {\bibinfo {author} {\bibfnamefont {Y.}~\bibnamefont
  {Zhou}}, \bibinfo {author} {\bibfnamefont {K.}~\bibnamefont {Kanoda}},\ and\
  \bibinfo {author} {\bibfnamefont {T.-K.}\ \bibnamefont {Ng}},\ }\bibfield
  {title} {\bibinfo {title} {Quantum spin liquid states},\ }\href
  {https://doi.org/10.1103/RevModPhys.89.025003} {\bibfield  {journal}
  {\bibinfo  {journal} {Rev. Mod. Phys.}\ }\textbf {\bibinfo {volume} {89}},\
  \bibinfo {pages} {025003} (\bibinfo {year} {2017})}\BibitemShut {NoStop}%
\bibitem [{\citenamefont {{Balents}}(2010)}]{Balents2010}%
  \BibitemOpen
  \bibfield  {author} {\bibinfo {author} {\bibfnamefont {L.}~\bibnamefont
  {{Balents}}},\ }\bibfield  {title} {\bibinfo {title} {{Spin liquids in
  frustrated magnets}},\ }\href {https://doi.org/10.1038/nature08917}
  {\bibfield  {journal} {\bibinfo  {journal} {Nature}\ }\textbf {\bibinfo
  {volume} {464}},\ \bibinfo {pages} {199} (\bibinfo {year}
  {2010})}\BibitemShut {NoStop}%
\bibitem [{\citenamefont {Shimizu}\ \emph {et~al.}(2003)\citenamefont
  {Shimizu}, \citenamefont {Miyagawa}, \citenamefont {Kanoda}, \citenamefont
  {Maesato},\ and\ \citenamefont {Saito}}]{Shimizu2003}%
  \BibitemOpen
  \bibfield  {author} {\bibinfo {author} {\bibfnamefont {Y.}~\bibnamefont
  {Shimizu}}, \bibinfo {author} {\bibfnamefont {K.}~\bibnamefont {Miyagawa}},
  \bibinfo {author} {\bibfnamefont {K.}~\bibnamefont {Kanoda}}, \bibinfo
  {author} {\bibfnamefont {M.}~\bibnamefont {Maesato}},\ and\ \bibinfo {author}
  {\bibfnamefont {G.}~\bibnamefont {Saito}},\ }\bibfield  {title} {\bibinfo
  {title} {Spin liquid state in an organic {Mott} insulator with a triangular
  lattice},\ }\href {https://doi.org/10.1103/PhysRevLett.91.107001} {\bibfield
  {journal} {\bibinfo  {journal} {Phys. Rev. Lett.}\ }\textbf {\bibinfo
  {volume} {91}},\ \bibinfo {pages} {107001} (\bibinfo {year}
  {2003})}\BibitemShut {NoStop}%
\bibitem [{\citenamefont {Yamashita}\ \emph {et~al.}(2010)\citenamefont
  {Yamashita}, \citenamefont {Nakata}, \citenamefont {Senshu}, \citenamefont
  {Nagata}, \citenamefont {Yamamoto}, \citenamefont {Kato}, \citenamefont
  {Shibauchi},\ and\ \citenamefont {Matsuda}}]{Yamashita2010}%
  \BibitemOpen
  \bibfield  {author} {\bibinfo {author} {\bibfnamefont {M.}~\bibnamefont
  {Yamashita}}, \bibinfo {author} {\bibfnamefont {N.}~\bibnamefont {Nakata}},
  \bibinfo {author} {\bibfnamefont {Y.}~\bibnamefont {Senshu}}, \bibinfo
  {author} {\bibfnamefont {M.}~\bibnamefont {Nagata}}, \bibinfo {author}
  {\bibfnamefont {H.~M.}\ \bibnamefont {Yamamoto}}, \bibinfo {author}
  {\bibfnamefont {R.}~\bibnamefont {Kato}}, \bibinfo {author} {\bibfnamefont
  {T.}~\bibnamefont {Shibauchi}},\ and\ \bibinfo {author} {\bibfnamefont
  {Y.}~\bibnamefont {Matsuda}},\ }\bibfield  {title} {\bibinfo {title} {Highly
  mobile gapless excitations in a two-dimensional candidate quantum spin
  liquid},\ }\href {https://doi.org/10.1126/science.1188200} {\bibfield
  {journal} {\bibinfo  {journal} {Science}\ }\textbf {\bibinfo {volume}
  {328}},\ \bibinfo {pages} {1246} (\bibinfo {year} {2010})}\BibitemShut
  {NoStop}%
\bibitem [{\citenamefont {Kanoda}\ and\ \citenamefont
  {Kato}(2011)}]{Kanoda2011}%
  \BibitemOpen
  \bibfield  {author} {\bibinfo {author} {\bibfnamefont {K.}~\bibnamefont
  {Kanoda}}\ and\ \bibinfo {author} {\bibfnamefont {R.}~\bibnamefont {Kato}},\
  }\bibfield  {title} {\bibinfo {title} {Mott physics in organic conductors
  with triangular lattices},\ }\href
  {https://doi.org/10.1146/annurev-conmatphys-062910-140521} {\bibfield
  {journal} {\bibinfo  {journal} {Annu. Rev. Condens. Matter Phys.}\ }\textbf
  {\bibinfo {volume} {2}},\ \bibinfo {pages} {167} (\bibinfo {year}
  {2011})}\BibitemShut {NoStop}%
\bibitem [{\citenamefont {Li}\ \emph {et~al.}(2015{\natexlab{a}})\citenamefont
  {Li}, \citenamefont {Liao}, \citenamefont {Zhang}, \citenamefont {Li},
  \citenamefont {Jin}, \citenamefont {Ling}, \citenamefont {Zhang},
  \citenamefont {Zou}, \citenamefont {Pi}, \citenamefont {Yang}, \citenamefont
  {Wang}, \citenamefont {Wu},\ and\ \citenamefont {Zhang}}]{Li2015YMGO1}%
  \BibitemOpen
  \bibfield  {author} {\bibinfo {author} {\bibfnamefont {Y.}~\bibnamefont
  {Li}}, \bibinfo {author} {\bibfnamefont {H.}~\bibnamefont {Liao}}, \bibinfo
  {author} {\bibfnamefont {Z.}~\bibnamefont {Zhang}}, \bibinfo {author}
  {\bibfnamefont {S.}~\bibnamefont {Li}}, \bibinfo {author} {\bibfnamefont
  {F.}~\bibnamefont {Jin}}, \bibinfo {author} {\bibfnamefont {L.}~\bibnamefont
  {Ling}}, \bibinfo {author} {\bibfnamefont {L.}~\bibnamefont {Zhang}},
  \bibinfo {author} {\bibfnamefont {Y.}~\bibnamefont {Zou}}, \bibinfo {author}
  {\bibfnamefont {L.}~\bibnamefont {Pi}}, \bibinfo {author} {\bibfnamefont
  {Z.}~\bibnamefont {Yang}}, \bibinfo {author} {\bibfnamefont {J.}~\bibnamefont
  {Wang}}, \bibinfo {author} {\bibfnamefont {Z.}~\bibnamefont {Wu}},\ and\
  \bibinfo {author} {\bibfnamefont {Q.}~\bibnamefont {Zhang}},\ }\bibfield
  {title} {\bibinfo {title} {Gapless quantum spin liquid ground state in the
  two-dimensional spin-1/2 triangular antiferromagnet {YbMgGaO$_4$}},\ }\href
  {https://doi.org/10.1038/srep16419} {\bibfield  {journal} {\bibinfo
  {journal} {Sci. Rep.}\ }\textbf {\bibinfo {volume} {5}},\ \bibinfo {pages}
  {16419} (\bibinfo {year} {2015}{\natexlab{a}})}\BibitemShut {NoStop}%
\bibitem [{\citenamefont {Li}\ \emph {et~al.}(2015{\natexlab{b}})\citenamefont
  {Li}, \citenamefont {Chen}, \citenamefont {Tong}, \citenamefont {Pi},
  \citenamefont {Liu}, \citenamefont {Yang}, \citenamefont {Wang},\ and\
  \citenamefont {Zhang}}]{Li2015YMGO2}%
  \BibitemOpen
  \bibfield  {author} {\bibinfo {author} {\bibfnamefont {Y.}~\bibnamefont
  {Li}}, \bibinfo {author} {\bibfnamefont {G.}~\bibnamefont {Chen}}, \bibinfo
  {author} {\bibfnamefont {W.}~\bibnamefont {Tong}}, \bibinfo {author}
  {\bibfnamefont {L.}~\bibnamefont {Pi}}, \bibinfo {author} {\bibfnamefont
  {J.}~\bibnamefont {Liu}}, \bibinfo {author} {\bibfnamefont {Z.}~\bibnamefont
  {Yang}}, \bibinfo {author} {\bibfnamefont {X.}~\bibnamefont {Wang}},\ and\
  \bibinfo {author} {\bibfnamefont {Q.}~\bibnamefont {Zhang}},\ }\bibfield
  {title} {\bibinfo {title} {Rare-earth triangular lattice spin liquid: A
  single-crystal study of {YbMgGaO$_4$}},\ }\href
  {https://doi.org/10.1103/PhysRevLett.115.167203} {\bibfield  {journal}
  {\bibinfo  {journal} {Phys. Rev. Lett.}\ }\textbf {\bibinfo {volume} {115}},\
  \bibinfo {pages} {167203} (\bibinfo {year} {2015}{\natexlab{b}})}\BibitemShut
  {NoStop}%
\bibitem [{\citenamefont {Shen}\ \emph {et~al.}(2016)\citenamefont {Shen},
  \citenamefont {Li}, \citenamefont {Wo}, \citenamefont {Li}, \citenamefont
  {Shen}, \citenamefont {Pan}, \citenamefont {Wang}, \citenamefont {Walker},
  \citenamefont {Steffens}, \citenamefont {Boehm}, \citenamefont {Hao},
  \citenamefont {Quintero-Castro}, \citenamefont {Harriger}, \citenamefont
  {Frontzek}, \citenamefont {Hao}, \citenamefont {Meng}, \citenamefont {Zhang},
  \citenamefont {Chen},\ and\ \citenamefont {Zhao}}]{Shen2016}%
  \BibitemOpen
  \bibfield  {author} {\bibinfo {author} {\bibfnamefont {Y.}~\bibnamefont
  {Shen}}, \bibinfo {author} {\bibfnamefont {Y.-D.}\ \bibnamefont {Li}},
  \bibinfo {author} {\bibfnamefont {H.}~\bibnamefont {Wo}}, \bibinfo {author}
  {\bibfnamefont {Y.}~\bibnamefont {Li}}, \bibinfo {author} {\bibfnamefont
  {S.}~\bibnamefont {Shen}}, \bibinfo {author} {\bibfnamefont {B.}~\bibnamefont
  {Pan}}, \bibinfo {author} {\bibfnamefont {Q.}~\bibnamefont {Wang}}, \bibinfo
  {author} {\bibfnamefont {H.~C.}\ \bibnamefont {Walker}}, \bibinfo {author}
  {\bibfnamefont {P.}~\bibnamefont {Steffens}}, \bibinfo {author}
  {\bibfnamefont {M.}~\bibnamefont {Boehm}}, \bibinfo {author} {\bibfnamefont
  {Y.}~\bibnamefont {Hao}}, \bibinfo {author} {\bibfnamefont {D.~L.}\
  \bibnamefont {Quintero-Castro}}, \bibinfo {author} {\bibfnamefont {L.~W.}\
  \bibnamefont {Harriger}}, \bibinfo {author} {\bibfnamefont {M.~D.}\
  \bibnamefont {Frontzek}}, \bibinfo {author} {\bibfnamefont {L.}~\bibnamefont
  {Hao}}, \bibinfo {author} {\bibfnamefont {S.}~\bibnamefont {Meng}}, \bibinfo
  {author} {\bibfnamefont {Q.}~\bibnamefont {Zhang}}, \bibinfo {author}
  {\bibfnamefont {G.}~\bibnamefont {Chen}},\ and\ \bibinfo {author}
  {\bibfnamefont {J.}~\bibnamefont {Zhao}},\ }\bibfield  {title} {\bibinfo
  {title} {Evidence for a spinon fermi surface in a triangular-lattice
  quantum-spin-liquid candidate},\ }\href {https://doi.org/10.1038/nature20614}
  {\bibfield  {journal} {\bibinfo  {journal} {Nature}\ }\textbf {\bibinfo
  {volume} {540}},\ \bibinfo {pages} {559} (\bibinfo {year}
  {2016})}\BibitemShut {NoStop}%
\bibitem [{\citenamefont {Paddison}\ \emph {et~al.}(2017)\citenamefont
  {Paddison}, \citenamefont {Daum}, \citenamefont {Dun}, \citenamefont
  {Ehlers}, \citenamefont {Liu}, \citenamefont {Stone}, \citenamefont {Zhou},\
  and\ \citenamefont {Mourigal}}]{Paddison2017}%
  \BibitemOpen
  \bibfield  {author} {\bibinfo {author} {\bibfnamefont {J.~A.~M.}\
  \bibnamefont {Paddison}}, \bibinfo {author} {\bibfnamefont {M.}~\bibnamefont
  {Daum}}, \bibinfo {author} {\bibfnamefont {Z.}~\bibnamefont {Dun}}, \bibinfo
  {author} {\bibfnamefont {G.}~\bibnamefont {Ehlers}}, \bibinfo {author}
  {\bibfnamefont {Y.}~\bibnamefont {Liu}}, \bibinfo {author} {\bibfnamefont
  {M.~B.}\ \bibnamefont {Stone}}, \bibinfo {author} {\bibfnamefont
  {H.}~\bibnamefont {Zhou}},\ and\ \bibinfo {author} {\bibfnamefont
  {M.}~\bibnamefont {Mourigal}},\ }\bibfield  {title} {\bibinfo {title}
  {Continuous excitations of the triangular-lattice quantum spin liquid
  {YbMgGaO$_4$}},\ }\href {https://doi.org/10.1038/nphys3971} {\bibfield
  {journal} {\bibinfo  {journal} {Nat. Phys.}\ }\textbf {\bibinfo {volume}
  {13}},\ \bibinfo {pages} {117} (\bibinfo {year} {2017})}\BibitemShut
  {NoStop}%
\bibitem [{\citenamefont {Shen}\ \emph {et~al.}(2018)\citenamefont {Shen},
  \citenamefont {Li}, \citenamefont {Walker}, \citenamefont {Steffens},
  \citenamefont {Boehm}, \citenamefont {Zhang}, \citenamefont {Shen},
  \citenamefont {Wo}, \citenamefont {Chen},\ and\ \citenamefont
  {Zhao}}]{Shen2018}%
  \BibitemOpen
  \bibfield  {author} {\bibinfo {author} {\bibfnamefont {Y.}~\bibnamefont
  {Shen}}, \bibinfo {author} {\bibfnamefont {Y.-D.}\ \bibnamefont {Li}},
  \bibinfo {author} {\bibfnamefont {H.~C.}\ \bibnamefont {Walker}}, \bibinfo
  {author} {\bibfnamefont {P.}~\bibnamefont {Steffens}}, \bibinfo {author}
  {\bibfnamefont {M.}~\bibnamefont {Boehm}}, \bibinfo {author} {\bibfnamefont
  {X.}~\bibnamefont {Zhang}}, \bibinfo {author} {\bibfnamefont
  {S.}~\bibnamefont {Shen}}, \bibinfo {author} {\bibfnamefont {H.}~\bibnamefont
  {Wo}}, \bibinfo {author} {\bibfnamefont {G.}~\bibnamefont {Chen}},\ and\
  \bibinfo {author} {\bibfnamefont {J.}~\bibnamefont {Zhao}},\ }\bibfield
  {title} {\bibinfo {title} {Fractionalized excitations in the partially
  magnetized spin liquid candidate {YbMgGaO$_4$}},\ }\href
  {https://doi.org/10.1038/s41467-018-06588-1} {\bibfield  {journal} {\bibinfo
  {journal} {Nat. Commun.}\ }\textbf {\bibinfo {volume} {9}},\ \bibinfo {pages}
  {4138} (\bibinfo {year} {2018})}\BibitemShut {NoStop}%
\bibitem [{\citenamefont {Liu}\ \emph {et~al.}(2018)\citenamefont {Liu},
  \citenamefont {Zhang}, \citenamefont {Ji}, \citenamefont {Liu}, \citenamefont
  {Li}, \citenamefont {Wang}, \citenamefont {Lei}, \citenamefont {Chen},\ and\
  \citenamefont {Zhang}}]{Liu2018}%
  \BibitemOpen
  \bibfield  {author} {\bibinfo {author} {\bibfnamefont {W.}~\bibnamefont
  {Liu}}, \bibinfo {author} {\bibfnamefont {Z.}~\bibnamefont {Zhang}}, \bibinfo
  {author} {\bibfnamefont {J.}~\bibnamefont {Ji}}, \bibinfo {author}
  {\bibfnamefont {Y.}~\bibnamefont {Liu}}, \bibinfo {author} {\bibfnamefont
  {J.}~\bibnamefont {Li}}, \bibinfo {author} {\bibfnamefont {X.}~\bibnamefont
  {Wang}}, \bibinfo {author} {\bibfnamefont {H.}~\bibnamefont {Lei}}, \bibinfo
  {author} {\bibfnamefont {G.}~\bibnamefont {Chen}},\ and\ \bibinfo {author}
  {\bibfnamefont {Q.}~\bibnamefont {Zhang}},\ }\bibfield  {title} {\bibinfo
  {title} {Rare-earth chalcogenides: A large family of triangular lattice spin
  liquid candidates},\ }\href {https://doi.org/10.1088/0256-307x/35/11/117501}
  {\bibfield  {journal} {\bibinfo  {journal} {Chin. Phys. Lett.}\ }\textbf
  {\bibinfo {volume} {35}},\ \bibinfo {pages} {117501} (\bibinfo {year}
  {2018})}\BibitemShut {NoStop}%
\bibitem [{\citenamefont {Bordelon}\ \emph {et~al.}(2019)\citenamefont
  {Bordelon}, \citenamefont {Kenney}, \citenamefont {Liu}, \citenamefont
  {Hogan}, \citenamefont {Posthuma}, \citenamefont {Kavand}, \citenamefont
  {Lyu}, \citenamefont {Sherwin}, \citenamefont {Butch}, \citenamefont {Brown},
  \citenamefont {Graf}, \citenamefont {Balents},\ and\ \citenamefont
  {Wilson}}]{Bordelon2019}%
  \BibitemOpen
  \bibfield  {author} {\bibinfo {author} {\bibfnamefont {M.~M.}\ \bibnamefont
  {Bordelon}}, \bibinfo {author} {\bibfnamefont {E.}~\bibnamefont {Kenney}},
  \bibinfo {author} {\bibfnamefont {C.}~\bibnamefont {Liu}}, \bibinfo {author}
  {\bibfnamefont {T.}~\bibnamefont {Hogan}}, \bibinfo {author} {\bibfnamefont
  {L.}~\bibnamefont {Posthuma}}, \bibinfo {author} {\bibfnamefont
  {M.}~\bibnamefont {Kavand}}, \bibinfo {author} {\bibfnamefont
  {Y.}~\bibnamefont {Lyu}}, \bibinfo {author} {\bibfnamefont {M.}~\bibnamefont
  {Sherwin}}, \bibinfo {author} {\bibfnamefont {N.~P.}\ \bibnamefont {Butch}},
  \bibinfo {author} {\bibfnamefont {C.}~\bibnamefont {Brown}}, \bibinfo
  {author} {\bibfnamefont {M.~J.}\ \bibnamefont {Graf}}, \bibinfo {author}
  {\bibfnamefont {L.}~\bibnamefont {Balents}},\ and\ \bibinfo {author}
  {\bibfnamefont {S.~D.}\ \bibnamefont {Wilson}},\ }\bibfield  {title}
  {\bibinfo {title} {Field-tunable quantum disordered ground state in the
  triangular-lattice antiferromagnet {NaYbO$_2$}},\ }\href
  {https://doi.org/10.1038/s41567-019-0594-5} {\bibfield  {journal} {\bibinfo
  {journal} {Nat. Phys.}\ }\textbf {\bibinfo {volume} {15}},\ \bibinfo {pages}
  {1058} (\bibinfo {year} {2019})}\BibitemShut {NoStop}%
\bibitem [{\citenamefont {Dai}\ \emph {et~al.}(2021)\citenamefont {Dai},
  \citenamefont {Zhang}, \citenamefont {Xie}, \citenamefont {Duan},
  \citenamefont {Gao}, \citenamefont {Zhu}, \citenamefont {Feng}, \citenamefont
  {Tao}, \citenamefont {Huang}, \citenamefont {Cao}, \citenamefont
  {Podlesnyak}, \citenamefont {Granroth}, \citenamefont {Everett},
  \citenamefont {Neuefeind}, \citenamefont {Voneshen}, \citenamefont {Wang},
  \citenamefont {Tan}, \citenamefont {Morosan}, \citenamefont {Wang},
  \citenamefont {Lin}, \citenamefont {Shu}, \citenamefont {Chen}, \citenamefont
  {Guo}, \citenamefont {Lu},\ and\ \citenamefont {Dai}}]{Dai2021}%
  \BibitemOpen
  \bibfield  {author} {\bibinfo {author} {\bibfnamefont {P.-L.}\ \bibnamefont
  {Dai}}, \bibinfo {author} {\bibfnamefont {G.}~\bibnamefont {Zhang}}, \bibinfo
  {author} {\bibfnamefont {Y.}~\bibnamefont {Xie}}, \bibinfo {author}
  {\bibfnamefont {C.}~\bibnamefont {Duan}}, \bibinfo {author} {\bibfnamefont
  {Y.}~\bibnamefont {Gao}}, \bibinfo {author} {\bibfnamefont {Z.}~\bibnamefont
  {Zhu}}, \bibinfo {author} {\bibfnamefont {E.}~\bibnamefont {Feng}}, \bibinfo
  {author} {\bibfnamefont {Z.}~\bibnamefont {Tao}}, \bibinfo {author}
  {\bibfnamefont {C.-L.}\ \bibnamefont {Huang}}, \bibinfo {author}
  {\bibfnamefont {H.}~\bibnamefont {Cao}}, \bibinfo {author} {\bibfnamefont
  {A.}~\bibnamefont {Podlesnyak}}, \bibinfo {author} {\bibfnamefont {G.~E.}\
  \bibnamefont {Granroth}}, \bibinfo {author} {\bibfnamefont {M.~S.}\
  \bibnamefont {Everett}}, \bibinfo {author} {\bibfnamefont {J.~C.}\
  \bibnamefont {Neuefeind}}, \bibinfo {author} {\bibfnamefont {D.}~\bibnamefont
  {Voneshen}}, \bibinfo {author} {\bibfnamefont {S.}~\bibnamefont {Wang}},
  \bibinfo {author} {\bibfnamefont {G.}~\bibnamefont {Tan}}, \bibinfo {author}
  {\bibfnamefont {E.}~\bibnamefont {Morosan}}, \bibinfo {author} {\bibfnamefont
  {X.}~\bibnamefont {Wang}}, \bibinfo {author} {\bibfnamefont {H.-Q.}\
  \bibnamefont {Lin}}, \bibinfo {author} {\bibfnamefont {L.}~\bibnamefont
  {Shu}}, \bibinfo {author} {\bibfnamefont {G.}~\bibnamefont {Chen}}, \bibinfo
  {author} {\bibfnamefont {Y.}~\bibnamefont {Guo}}, \bibinfo {author}
  {\bibfnamefont {X.}~\bibnamefont {Lu}},\ and\ \bibinfo {author}
  {\bibfnamefont {P.}~\bibnamefont {Dai}},\ }\bibfield  {title} {\bibinfo
  {title} {Spinon fermi surface spin liquid in a triangular lattice
  antiferromagnet {{NaYbSe}$_{2}$}},\ }\href
  {https://doi.org/10.1103/PhysRevX.11.021044} {\bibfield  {journal} {\bibinfo
  {journal} {Phys. Rev. X}\ }\textbf {\bibinfo {volume} {11}},\ \bibinfo
  {pages} {021044} (\bibinfo {year} {2021})}\BibitemShut {NoStop}%
\bibitem [{\citenamefont {Zhong}\ \emph {et~al.}(2019)\citenamefont {Zhong},
  \citenamefont {Guo}, \citenamefont {Xu}, \citenamefont {Xu},\ and\
  \citenamefont {Cava}}]{Zhong2019}%
  \BibitemOpen
  \bibfield  {author} {\bibinfo {author} {\bibfnamefont {R.}~\bibnamefont
  {Zhong}}, \bibinfo {author} {\bibfnamefont {S.}~\bibnamefont {Guo}}, \bibinfo
  {author} {\bibfnamefont {G.}~\bibnamefont {Xu}}, \bibinfo {author}
  {\bibfnamefont {Z.}~\bibnamefont {Xu}},\ and\ \bibinfo {author}
  {\bibfnamefont {R.~J.}\ \bibnamefont {Cava}},\ }\bibfield  {title} {\bibinfo
  {title} {Strong quantum fluctuations in a quantum spin liquid candidate with
  a {Co}-based triangular lattice},\ }\href
  {https://doi.org/10.1073/pnas.1906483116} {\bibfield  {journal} {\bibinfo
  {journal} {Proc. Natl. Acad. Sci. U.S.A.}\ }\textbf {\bibinfo {volume}
  {116}},\ \bibinfo {pages} {14505} (\bibinfo {year} {2019})}\BibitemShut
  {NoStop}%
\bibitem [{\citenamefont {Li}\ \emph {et~al.}(2020{\natexlab{a}})\citenamefont
  {Li}, \citenamefont {Huang}, \citenamefont {Yue}, \citenamefont {Chu},
  \citenamefont {Chen}, \citenamefont {Choi}, \citenamefont {Zhao},
  \citenamefont {Zhou},\ and\ \citenamefont {Sun}}]{LiN2020}%
  \BibitemOpen
  \bibfield  {author} {\bibinfo {author} {\bibfnamefont {N.}~\bibnamefont
  {Li}}, \bibinfo {author} {\bibfnamefont {Q.}~\bibnamefont {Huang}}, \bibinfo
  {author} {\bibfnamefont {X.~Y.}\ \bibnamefont {Yue}}, \bibinfo {author}
  {\bibfnamefont {W.~J.}\ \bibnamefont {Chu}}, \bibinfo {author} {\bibfnamefont
  {Q.}~\bibnamefont {Chen}}, \bibinfo {author} {\bibfnamefont {E.~S.}\
  \bibnamefont {Choi}}, \bibinfo {author} {\bibfnamefont {X.}~\bibnamefont
  {Zhao}}, \bibinfo {author} {\bibfnamefont {H.~D.}\ \bibnamefont {Zhou}},\
  and\ \bibinfo {author} {\bibfnamefont {X.~F.}\ \bibnamefont {Sun}},\
  }\bibfield  {title} {\bibinfo {title} {{Possible itinerant excitations and
  quantum spin state transitions in the effective spin-1/2 triangular-lattice
  antiferromagnet Na$_2$BaCo(PO$_4$)$_2$}},\ }\href
  {https://doi.org/10.1038/s41467-020-18041-3} {\bibfield  {journal} {\bibinfo
  {journal} {Nat. Commun.}\ }\textbf {\bibinfo {volume} {11}},\ \bibinfo
  {pages} {4216} (\bibinfo {year} {2020}{\natexlab{a}})}\BibitemShut {NoStop}%
\bibitem [{\citenamefont {Lee}\ \emph {et~al.}(2021)\citenamefont {Lee},
  \citenamefont {Lee}, \citenamefont {Berlie}, \citenamefont {Hillier},
  \citenamefont {Adroja}, \citenamefont {Zhong}, \citenamefont {Cava},
  \citenamefont {Jang},\ and\ \citenamefont {Choi}}]{Lee2021}%
  \BibitemOpen
  \bibfield  {author} {\bibinfo {author} {\bibfnamefont {S.}~\bibnamefont
  {Lee}}, \bibinfo {author} {\bibfnamefont {C.~H.}\ \bibnamefont {Lee}},
  \bibinfo {author} {\bibfnamefont {A.}~\bibnamefont {Berlie}}, \bibinfo
  {author} {\bibfnamefont {A.~D.}\ \bibnamefont {Hillier}}, \bibinfo {author}
  {\bibfnamefont {D.~T.}\ \bibnamefont {Adroja}}, \bibinfo {author}
  {\bibfnamefont {R.}~\bibnamefont {Zhong}}, \bibinfo {author} {\bibfnamefont
  {R.~J.}\ \bibnamefont {Cava}}, \bibinfo {author} {\bibfnamefont {Z.~H.}\
  \bibnamefont {Jang}},\ and\ \bibinfo {author} {\bibfnamefont {K.-Y.}\
  \bibnamefont {Choi}},\ }\bibfield  {title} {\bibinfo {title} {{Temporal and
  field evolution of spin excitations in the disorder-free triangular
  antiferromagnet Na$_2$BaCo(PO$_4$)$_2$}},\ }\href
  {https://doi.org/10.1103/PhysRevB.103.024413} {\bibfield  {journal} {\bibinfo
   {journal} {Phys. Rev. B}\ }\textbf {\bibinfo {volume} {103}},\ \bibinfo
  {pages} {024413} (\bibinfo {year} {2021})}\BibitemShut {NoStop}%
\bibitem [{\citenamefont {Wellm}\ \emph {et~al.}(2021)\citenamefont {Wellm},
  \citenamefont {Roscher}, \citenamefont {Zeisner}, \citenamefont {Alfonsov},
  \citenamefont {Zhong}, \citenamefont {Cava}, \citenamefont {Savoyant},
  \citenamefont {Hayn}, \citenamefont {van~den Brink}, \citenamefont
  {B\"uchner}, \citenamefont {Janson},\ and\ \citenamefont
  {Kataev}}]{Wellm2021}%
  \BibitemOpen
  \bibfield  {author} {\bibinfo {author} {\bibfnamefont {C.}~\bibnamefont
  {Wellm}}, \bibinfo {author} {\bibfnamefont {W.}~\bibnamefont {Roscher}},
  \bibinfo {author} {\bibfnamefont {J.}~\bibnamefont {Zeisner}}, \bibinfo
  {author} {\bibfnamefont {A.}~\bibnamefont {Alfonsov}}, \bibinfo {author}
  {\bibfnamefont {R.}~\bibnamefont {Zhong}}, \bibinfo {author} {\bibfnamefont
  {R.~J.}\ \bibnamefont {Cava}}, \bibinfo {author} {\bibfnamefont
  {A.}~\bibnamefont {Savoyant}}, \bibinfo {author} {\bibfnamefont
  {R.}~\bibnamefont {Hayn}}, \bibinfo {author} {\bibfnamefont {J.}~\bibnamefont
  {van~den Brink}}, \bibinfo {author} {\bibfnamefont {B.}~\bibnamefont
  {B\"uchner}}, \bibinfo {author} {\bibfnamefont {O.}~\bibnamefont {Janson}},\
  and\ \bibinfo {author} {\bibfnamefont {V.}~\bibnamefont {Kataev}},\
  }\bibfield  {title} {\bibinfo {title} {Frustration enhanced by {Kitaev}
  exchange in a $j_{\text{eff}}=\frac{1}{2}$ triangular antiferromagnet},\
  }\href {https://doi.org/10.1103/PhysRevB.104.L100420} {\bibfield  {journal}
  {\bibinfo  {journal} {Phys. Rev. B}\ }\textbf {\bibinfo {volume} {104}},\
  \bibinfo {pages} {L100420} (\bibinfo {year} {2021})}\BibitemShut {NoStop}%
\bibitem [{\citenamefont {Gao}\ \emph {et~al.}(2022)\citenamefont {Gao},
  \citenamefont {Fan}, \citenamefont {Li}, \citenamefont {Yang}, \citenamefont
  {Zeng}, \citenamefont {Sheng}, \citenamefont {Zhong}, \citenamefont {Qi},
  \citenamefont {Wan},\ and\ \citenamefont {Li}}]{Gao2022QMats}%
  \BibitemOpen
  \bibfield  {author} {\bibinfo {author} {\bibfnamefont {Y.}~\bibnamefont
  {Gao}}, \bibinfo {author} {\bibfnamefont {Y.-C.}\ \bibnamefont {Fan}},
  \bibinfo {author} {\bibfnamefont {H.}~\bibnamefont {Li}}, \bibinfo {author}
  {\bibfnamefont {F.}~\bibnamefont {Yang}}, \bibinfo {author} {\bibfnamefont
  {X.-T.}\ \bibnamefont {Zeng}}, \bibinfo {author} {\bibfnamefont {X.-L.}\
  \bibnamefont {Sheng}}, \bibinfo {author} {\bibfnamefont {R.}~\bibnamefont
  {Zhong}}, \bibinfo {author} {\bibfnamefont {Y.}~\bibnamefont {Qi}}, \bibinfo
  {author} {\bibfnamefont {Y.}~\bibnamefont {Wan}},\ and\ \bibinfo {author}
  {\bibfnamefont {W.}~\bibnamefont {Li}},\ }\bibfield  {title} {\bibinfo
  {title} {Spin supersolidity in nearly ideal easy-axis triangular quantum
  antiferromagnet {Na$_2$BaCo(PO$_4$)$_2$}},\ }\href
  {https://doi.org/10.1038/s41535-022-00500-3} {\bibfield  {journal} {\bibinfo
  {journal} {npj Quantum Mater.}\ }\textbf {\bibinfo {volume} {7}},\ \bibinfo
  {pages} {89} (\bibinfo {year} {2022})}\BibitemShut {NoStop}%
\bibitem [{\citenamefont {Cevallos}\ \emph {et~al.}(2018)\citenamefont
  {Cevallos}, \citenamefont {Stolze}, \citenamefont {Kong},\ and\ \citenamefont
  {Cava}}]{Cava2018}%
  \BibitemOpen
  \bibfield  {author} {\bibinfo {author} {\bibfnamefont {F.~A.}\ \bibnamefont
  {Cevallos}}, \bibinfo {author} {\bibfnamefont {K.}~\bibnamefont {Stolze}},
  \bibinfo {author} {\bibfnamefont {T.}~\bibnamefont {Kong}},\ and\ \bibinfo
  {author} {\bibfnamefont {R.~J.}\ \bibnamefont {Cava}},\ }\bibfield  {title}
  {\bibinfo {title} {Anisotropic magnetic properties of the triangular plane
  lattice material {TmMgGaO}$_4$},\ }\href
  {https://doi.org/https://doi.org/10.1016/j.materresbull.2018.04.042}
  {\bibfield  {journal} {\bibinfo  {journal} {Mater. Res. Bull.}\ }\textbf
  {\bibinfo {volume} {105}},\ \bibinfo {pages} {154} (\bibinfo {year}
  {2018})}\BibitemShut {NoStop}%
\bibitem [{\citenamefont {Shen}\ \emph {et~al.}(2019)\citenamefont {Shen},
  \citenamefont {Liu}, \citenamefont {Qin}, \citenamefont {Shen}, \citenamefont
  {Li}, \citenamefont {Bewley}, \citenamefont {Schneidewind}, \citenamefont
  {Chen},\ and\ \citenamefont {Zhao}}]{Shen2019}%
  \BibitemOpen
  \bibfield  {author} {\bibinfo {author} {\bibfnamefont {Y.}~\bibnamefont
  {Shen}}, \bibinfo {author} {\bibfnamefont {C.}~\bibnamefont {Liu}}, \bibinfo
  {author} {\bibfnamefont {Y.}~\bibnamefont {Qin}}, \bibinfo {author}
  {\bibfnamefont {S.}~\bibnamefont {Shen}}, \bibinfo {author} {\bibfnamefont
  {Y.-D.}\ \bibnamefont {Li}}, \bibinfo {author} {\bibfnamefont
  {R.}~\bibnamefont {Bewley}}, \bibinfo {author} {\bibfnamefont
  {A.}~\bibnamefont {Schneidewind}}, \bibinfo {author} {\bibfnamefont
  {G.}~\bibnamefont {Chen}},\ and\ \bibinfo {author} {\bibfnamefont
  {J.}~\bibnamefont {Zhao}},\ }\bibfield  {title} {\bibinfo {title}
  {Intertwined dipolar and multipolar order in the triangular-lattice magnet
  {TmMgGaO$_4$}},\ }\href {https://doi.org/10.1038/s41467-019-12410-3}
  {\bibfield  {journal} {\bibinfo  {journal} {Nat. Commun.}\ }\textbf {\bibinfo
  {volume} {10}},\ \bibinfo {pages} {4530} (\bibinfo {year}
  {2019})}\BibitemShut {NoStop}%
\bibitem [{\citenamefont {Li}\ \emph {et~al.}(2020{\natexlab{b}})\citenamefont
  {Li}, \citenamefont {Bachus}, \citenamefont {Deng}, \citenamefont {Schmidt},
  \citenamefont {Thoma}, \citenamefont {Hutanu}, \citenamefont {Tokiwa},
  \citenamefont {Tsirlin},\ and\ \citenamefont {Gegenwart}}]{Li2020}%
  \BibitemOpen
  \bibfield  {author} {\bibinfo {author} {\bibfnamefont {Y.}~\bibnamefont
  {Li}}, \bibinfo {author} {\bibfnamefont {S.}~\bibnamefont {Bachus}}, \bibinfo
  {author} {\bibfnamefont {H.}~\bibnamefont {Deng}}, \bibinfo {author}
  {\bibfnamefont {W.}~\bibnamefont {Schmidt}}, \bibinfo {author} {\bibfnamefont
  {H.}~\bibnamefont {Thoma}}, \bibinfo {author} {\bibfnamefont
  {V.}~\bibnamefont {Hutanu}}, \bibinfo {author} {\bibfnamefont
  {Y.}~\bibnamefont {Tokiwa}}, \bibinfo {author} {\bibfnamefont {A.~A.}\
  \bibnamefont {Tsirlin}},\ and\ \bibinfo {author} {\bibfnamefont
  {P.}~\bibnamefont {Gegenwart}},\ }\bibfield  {title} {\bibinfo {title}
  {Partial up-up-down order with the continuously distributed order parameter
  in the triangular antiferromagnet {TmMgGaO}$_4$},\ }\href
  {https://doi.org/10.1103/PhysRevX.10.011007} {\bibfield  {journal} {\bibinfo
  {journal} {Phys. Rev. X}\ }\textbf {\bibinfo {volume} {10}},\ \bibinfo
  {pages} {011007} (\bibinfo {year} {2020}{\natexlab{b}})}\BibitemShut
  {NoStop}%
\bibitem [{\citenamefont {Li}\ \emph {et~al.}(2020{\natexlab{c}})\citenamefont
  {Li}, \citenamefont {Liao}, \citenamefont {Chen}, \citenamefont {Zeng},
  \citenamefont {Sheng}, \citenamefont {Qi}, \citenamefont {Meng},\ and\
  \citenamefont {Li}}]{Lih2020}%
  \BibitemOpen
  \bibfield  {author} {\bibinfo {author} {\bibfnamefont {H.}~\bibnamefont
  {Li}}, \bibinfo {author} {\bibfnamefont {Y.~D.}\ \bibnamefont {Liao}},
  \bibinfo {author} {\bibfnamefont {B.-B.}\ \bibnamefont {Chen}}, \bibinfo
  {author} {\bibfnamefont {X.-T.}\ \bibnamefont {Zeng}}, \bibinfo {author}
  {\bibfnamefont {X.-L.}\ \bibnamefont {Sheng}}, \bibinfo {author}
  {\bibfnamefont {Y.}~\bibnamefont {Qi}}, \bibinfo {author} {\bibfnamefont
  {Z.~Y.}\ \bibnamefont {Meng}},\ and\ \bibinfo {author} {\bibfnamefont
  {W.}~\bibnamefont {Li}},\ }\bibfield  {title} {\bibinfo {title}
  {{Kosterlitz-Thouless} melting of magnetic order in the triangular quantum
  {Ising} material {TmMgGaO$_4$}},\ }\href
  {https://doi.org/10.1038/s41467-020-14907-8} {\bibfield  {journal} {\bibinfo
  {journal} {Nat. Commun.}\ }\textbf {\bibinfo {volume} {11}},\ \bibinfo
  {pages} {1111} (\bibinfo {year} {2020}{\natexlab{c}})}\BibitemShut {NoStop}%
\bibitem [{\citenamefont {Hu}\ \emph {et~al.}(2020)\citenamefont {Hu},
  \citenamefont {Ma}, \citenamefont {Liao}, \citenamefont {Li}, \citenamefont
  {Ma}, \citenamefont {Cui}, \citenamefont {Shangguan}, \citenamefont {Huang},
  \citenamefont {Qi}, \citenamefont {Li}, \citenamefont {Meng}, \citenamefont
  {Wen},\ and\ \citenamefont {Yu}}]{ZHu2020}%
  \BibitemOpen
  \bibfield  {author} {\bibinfo {author} {\bibfnamefont {Z.}~\bibnamefont
  {Hu}}, \bibinfo {author} {\bibfnamefont {Z.}~\bibnamefont {Ma}}, \bibinfo
  {author} {\bibfnamefont {Y.-D.}\ \bibnamefont {Liao}}, \bibinfo {author}
  {\bibfnamefont {H.}~\bibnamefont {Li}}, \bibinfo {author} {\bibfnamefont
  {C.}~\bibnamefont {Ma}}, \bibinfo {author} {\bibfnamefont {Y.}~\bibnamefont
  {Cui}}, \bibinfo {author} {\bibfnamefont {Y.}~\bibnamefont {Shangguan}},
  \bibinfo {author} {\bibfnamefont {Z.}~\bibnamefont {Huang}}, \bibinfo
  {author} {\bibfnamefont {Y.}~\bibnamefont {Qi}}, \bibinfo {author}
  {\bibfnamefont {W.}~\bibnamefont {Li}}, \bibinfo {author} {\bibfnamefont
  {Z.~Y.}\ \bibnamefont {Meng}}, \bibinfo {author} {\bibfnamefont
  {J.}~\bibnamefont {Wen}},\ and\ \bibinfo {author} {\bibfnamefont
  {W.}~\bibnamefont {Yu}},\ }\bibfield  {title} {\bibinfo {title} {Evidence of
  the {Berezinskii-Kosterlitz-Thouless} phase in a frustrated magnet},\ }\href
  {https://doi.org/10.1038/s41467-020-19380-x} {\bibfield  {journal} {\bibinfo
  {journal} {Nat. Commun.}\ }\textbf {\bibinfo {volume} {11}},\ \bibinfo
  {pages} {5631} (\bibinfo {year} {2020})}\BibitemShut {NoStop}%
\bibitem [{\citenamefont {Dun}\ \emph {et~al.}(2021)\citenamefont {Dun},
  \citenamefont {Daum}, \citenamefont {Baral}, \citenamefont {Fischer},
  \citenamefont {Cao}, \citenamefont {Liu}, \citenamefont {Stone},
  \citenamefont {Rodriguez-Rivera}, \citenamefont {Choi}, \citenamefont
  {Huang}, \citenamefont {Zhou}, \citenamefont {Mourigal},\ and\ \citenamefont
  {Frandsen}}]{Dun2020neutron}%
  \BibitemOpen
  \bibfield  {author} {\bibinfo {author} {\bibfnamefont {Z.}~\bibnamefont
  {Dun}}, \bibinfo {author} {\bibfnamefont {M.}~\bibnamefont {Daum}}, \bibinfo
  {author} {\bibfnamefont {R.}~\bibnamefont {Baral}}, \bibinfo {author}
  {\bibfnamefont {H.~E.}\ \bibnamefont {Fischer}}, \bibinfo {author}
  {\bibfnamefont {H.}~\bibnamefont {Cao}}, \bibinfo {author} {\bibfnamefont
  {Y.}~\bibnamefont {Liu}}, \bibinfo {author} {\bibfnamefont {M.~B.}\
  \bibnamefont {Stone}}, \bibinfo {author} {\bibfnamefont {J.~A.}\ \bibnamefont
  {Rodriguez-Rivera}}, \bibinfo {author} {\bibfnamefont {E.~S.}\ \bibnamefont
  {Choi}}, \bibinfo {author} {\bibfnamefont {Q.}~\bibnamefont {Huang}},
  \bibinfo {author} {\bibfnamefont {H.}~\bibnamefont {Zhou}}, \bibinfo {author}
  {\bibfnamefont {M.}~\bibnamefont {Mourigal}},\ and\ \bibinfo {author}
  {\bibfnamefont {B.~A.}\ \bibnamefont {Frandsen}},\ }\bibfield  {title}
  {\bibinfo {title} {Neutron scattering investigation of proposed
  {Kosterlitz-Thouless} transitions in the triangular-lattice ising
  antiferromagnet {TmMgGaO}$_4$},\ }\href
  {https://doi.org/10.1103/PhysRevB.103.064424} {\bibfield  {journal} {\bibinfo
   {journal} {Phys. Rev. B}\ }\textbf {\bibinfo {volume} {103}},\ \bibinfo
  {pages} {064424} (\bibinfo {year} {2021})}\BibitemShut {NoStop}%
\bibitem [{SM()}]{SM}%
  \BibitemOpen
  \href@noop {} {\bibinfo {title} {{Supplementary Sec.~1 describes the KBGB
  sample preparation and their XRD charactherization. The thermodynamic and
  magnetocaloric measurements are elaborated in Sec.~2, and Sec.~3 is devoted
  to the HD model calculations.}}}\BibitemShut {Stop}%
\bibitem [{\citenamefont {Yao}\ \emph {et~al.}(2018)\citenamefont {Yao},
  \citenamefont {Zaletel}, \citenamefont {Stamper-Kurn},\ and\ \citenamefont
  {Vishwanath}}]{Yao2018}%
  \BibitemOpen
  \bibfield  {author} {\bibinfo {author} {\bibfnamefont {N.~Y.}\ \bibnamefont
  {Yao}}, \bibinfo {author} {\bibfnamefont {M.~P.}\ \bibnamefont {Zaletel}},
  \bibinfo {author} {\bibfnamefont {D.~M.}\ \bibnamefont {Stamper-Kurn}},\ and\
  \bibinfo {author} {\bibfnamefont {A.}~\bibnamefont {Vishwanath}},\ }\bibfield
   {title} {\bibinfo {title} {A quantum dipolar spin liquid},\ }\href
  {https://doi.org/10.1038/s41567-017-0030-7} {\bibfield  {journal} {\bibinfo
  {journal} {Nat. Phys.}\ }\textbf {\bibinfo {volume} {14}},\ \bibinfo {pages}
  {405} (\bibinfo {year} {2018})}\BibitemShut {NoStop}%
\bibitem [{\citenamefont {Zeng}\ \emph {et~al.}(2020)\citenamefont {Zeng},
  \citenamefont {Ma}, \citenamefont {Gao}, \citenamefont {Tian}, \citenamefont
  {Ling},\ and\ \citenamefont {Pi}}]{Zeng2020}%
  \BibitemOpen
  \bibfield  {author} {\bibinfo {author} {\bibfnamefont {K.~Y.}\ \bibnamefont
  {Zeng}}, \bibinfo {author} {\bibfnamefont {L.}~\bibnamefont {Ma}}, \bibinfo
  {author} {\bibfnamefont {Y.~X.}\ \bibnamefont {Gao}}, \bibinfo {author}
  {\bibfnamefont {Z.~M.}\ \bibnamefont {Tian}}, \bibinfo {author}
  {\bibfnamefont {L.~S.}\ \bibnamefont {Ling}},\ and\ \bibinfo {author}
  {\bibfnamefont {L.}~\bibnamefont {Pi}},\ }\bibfield  {title} {\bibinfo
  {title} {{NMR} study of the spin excitations in the frustrated
  antiferromagnet {Yb{(}BaBO$_{3}${)}$_{3}$} with a triangular lattice},\
  }\href {https://doi.org/10.1103/PhysRevB.102.045149} {\bibfield  {journal}
  {\bibinfo  {journal} {Phys. Rev. B}\ }\textbf {\bibinfo {volume} {102}},\
  \bibinfo {pages} {045149} (\bibinfo {year} {2020})}\BibitemShut {NoStop}%
\bibitem [{\citenamefont {Bag}\ \emph {et~al.}(2021)\citenamefont {Bag},
  \citenamefont {Ennis}, \citenamefont {Liu}, \citenamefont {Dissanayake},
  \citenamefont {Shi}, \citenamefont {Liu}, \citenamefont {Balents},\ and\
  \citenamefont {Haravifard}}]{Bag2021}%
  \BibitemOpen
  \bibfield  {author} {\bibinfo {author} {\bibfnamefont {R.}~\bibnamefont
  {Bag}}, \bibinfo {author} {\bibfnamefont {M.}~\bibnamefont {Ennis}}, \bibinfo
  {author} {\bibfnamefont {C.}~\bibnamefont {Liu}}, \bibinfo {author}
  {\bibfnamefont {S.~E.}\ \bibnamefont {Dissanayake}}, \bibinfo {author}
  {\bibfnamefont {Z.}~\bibnamefont {Shi}}, \bibinfo {author} {\bibfnamefont
  {J.}~\bibnamefont {Liu}}, \bibinfo {author} {\bibfnamefont {L.}~\bibnamefont
  {Balents}},\ and\ \bibinfo {author} {\bibfnamefont {S.}~\bibnamefont
  {Haravifard}},\ }\bibfield  {title} {\bibinfo {title} {Realization of quantum
  dipoles in triangular lattice crystal {Ba$_3$Yb(BO$_3$)$_3$}},\ }\href
  {https://doi.org/10.1103/PhysRevB.104.L220403} {\bibfield  {journal}
  {\bibinfo  {journal} {Phys. Rev. B}\ }\textbf {\bibinfo {volume} {104}},\
  \bibinfo {pages} {L220403} (\bibinfo {year} {2021})}\BibitemShut {NoStop}%
\bibitem [{\citenamefont {{Cho}}\ \emph {et~al.}(2021)\citenamefont {{Cho}},
  \citenamefont {{Blundell}}, \citenamefont {{Shiroka}}, \citenamefont
  {{MacFarquharson}}, \citenamefont {{Prabhakaran}},\ and\ \citenamefont
  {{Coldea}}}]{Cho2021BYBO}%
  \BibitemOpen
  \bibfield  {author} {\bibinfo {author} {\bibfnamefont {H.}~\bibnamefont
  {{Cho}}}, \bibinfo {author} {\bibfnamefont {S.~J.}\ \bibnamefont
  {{Blundell}}}, \bibinfo {author} {\bibfnamefont {T.}~\bibnamefont
  {{Shiroka}}}, \bibinfo {author} {\bibfnamefont {K.}~\bibnamefont
  {{MacFarquharson}}}, \bibinfo {author} {\bibfnamefont {D.}~\bibnamefont
  {{Prabhakaran}}},\ and\ \bibinfo {author} {\bibfnamefont {R.}~\bibnamefont
  {{Coldea}}},\ }\bibfield  {title} {\bibinfo {title} {{Studies on Novel
  Yb-based Candidate Triangular Quantum Antiferromagnets: {Ba$_3$YbB$_3$O$_9$}
  and Ba$_3$YbB$_9$O$_{18}$}},\ }\href {https://arxiv.org/abs/2104.01005}
  {\bibfield  {journal} {\bibinfo  {journal} {arXiv:2104.01005}\ } (\bibinfo
  {year} {2021})}\BibitemShut {NoStop}%
\bibitem [{\citenamefont {Khatua}\ \emph {et~al.}(2022)\citenamefont {Khatua},
  \citenamefont {Pregelj}, \citenamefont {Elghandour}, \citenamefont
  {Jagli\ifmmode~\check{c}\else \v{c}\fi{}ic}, \citenamefont {Klingeler},
  \citenamefont {Zorko},\ and\ \citenamefont {Khuntia}}]{Khatua2022Ba3RB9O18}%
  \BibitemOpen
  \bibfield  {author} {\bibinfo {author} {\bibfnamefont {J.}~\bibnamefont
  {Khatua}}, \bibinfo {author} {\bibfnamefont {M.}~\bibnamefont {Pregelj}},
  \bibinfo {author} {\bibfnamefont {A.}~\bibnamefont {Elghandour}}, \bibinfo
  {author} {\bibfnamefont {Z.}~\bibnamefont {Jagli\ifmmode~\check{c}\else
  \v{c}\fi{}ic}}, \bibinfo {author} {\bibfnamefont {R.}~\bibnamefont
  {Klingeler}}, \bibinfo {author} {\bibfnamefont {A.}~\bibnamefont {Zorko}},\
  and\ \bibinfo {author} {\bibfnamefont {P.}~\bibnamefont {Khuntia}},\
  }\bibfield  {title} {\bibinfo {title} {{Magnetic properties of triangular
  lattice antiferromagnets {Ba$_3$RB$_9$O$_{18}$ (R = Yb, Er)}}},\ }\href
  {https://doi.org/10.1103/PhysRevB.106.104408} {\bibfield  {journal} {\bibinfo
   {journal} {Phys. Rev. B}\ }\textbf {\bibinfo {volume} {106}},\ \bibinfo
  {pages} {104408} (\bibinfo {year} {2022})}\BibitemShut {NoStop}%
\bibitem [{\citenamefont {Jiang}\ \emph {et~al.}(2022)\citenamefont {Jiang},
  \citenamefont {Yang}, \citenamefont {Gao}, \citenamefont {Wan}, \citenamefont
  {Zhu}, \citenamefont {Shiroka}, \citenamefont {Chen}, \citenamefont {Wu},
  \citenamefont {Li}, \citenamefont {Jiao}, \citenamefont {Chen}, \citenamefont
  {Bao}, \citenamefont {Tian},\ and\ \citenamefont {Shu}}]{Jiang2022}%
  \BibitemOpen
  \bibfield  {author} {\bibinfo {author} {\bibfnamefont {C.~Y.}\ \bibnamefont
  {Jiang}}, \bibinfo {author} {\bibfnamefont {Y.~X.}\ \bibnamefont {Yang}},
  \bibinfo {author} {\bibfnamefont {Y.~X.}\ \bibnamefont {Gao}}, \bibinfo
  {author} {\bibfnamefont {Z.~T.}\ \bibnamefont {Wan}}, \bibinfo {author}
  {\bibfnamefont {Z.~H.}\ \bibnamefont {Zhu}}, \bibinfo {author} {\bibfnamefont
  {T.}~\bibnamefont {Shiroka}}, \bibinfo {author} {\bibfnamefont {C.~S.}\
  \bibnamefont {Chen}}, \bibinfo {author} {\bibfnamefont {Q.}~\bibnamefont
  {Wu}}, \bibinfo {author} {\bibfnamefont {X.}~\bibnamefont {Li}}, \bibinfo
  {author} {\bibfnamefont {J.~C.}\ \bibnamefont {Jiao}}, \bibinfo {author}
  {\bibfnamefont {K.~W.}\ \bibnamefont {Chen}}, \bibinfo {author}
  {\bibfnamefont {Y.}~\bibnamefont {Bao}}, \bibinfo {author} {\bibfnamefont
  {Z.~M.}\ \bibnamefont {Tian}},\ and\ \bibinfo {author} {\bibfnamefont
  {L.}~\bibnamefont {Shu}},\ }\bibfield  {title} {\bibinfo {title} {Spin
  excitations in the quantum dipolar magnet {Yb(BaBO$_3$)$_3$}},\ }\href
  {https://doi.org/10.1103/PhysRevB.106.014409} {\bibfield  {journal} {\bibinfo
   {journal} {Phys. Rev. B}\ }\textbf {\bibinfo {volume} {106}},\ \bibinfo
  {pages} {014409} (\bibinfo {year} {2022})}\BibitemShut {NoStop}%
\bibitem [{\citenamefont {Liu}\ \emph {et~al.}(2022)\citenamefont {Liu},
  \citenamefont {Gao}, \citenamefont {Li}, \citenamefont {Jin}, \citenamefont
  {Xiang}, \citenamefont {Jin}, \citenamefont {Chen}, \citenamefont {Li},\ and\
  \citenamefont {Su}}]{Liu2022}%
  \BibitemOpen
  \bibfield  {author} {\bibinfo {author} {\bibfnamefont {X.-Y.}\ \bibnamefont
  {Liu}}, \bibinfo {author} {\bibfnamefont {Y.}~\bibnamefont {Gao}}, \bibinfo
  {author} {\bibfnamefont {H.}~\bibnamefont {Li}}, \bibinfo {author}
  {\bibfnamefont {W.}~\bibnamefont {Jin}}, \bibinfo {author} {\bibfnamefont
  {J.}~\bibnamefont {Xiang}}, \bibinfo {author} {\bibfnamefont
  {H.}~\bibnamefont {Jin}}, \bibinfo {author} {\bibfnamefont {Z.}~\bibnamefont
  {Chen}}, \bibinfo {author} {\bibfnamefont {W.}~\bibnamefont {Li}},\ and\
  \bibinfo {author} {\bibfnamefont {G.}~\bibnamefont {Su}},\ }\bibfield
  {title} {\bibinfo {title} {Quantum spin liquid candidate as superior
  refrigerant in cascade demagnetization cooling},\ }\href
  {https://doi.org/10.1038/s42005-022-01010-1} {\bibfield  {journal} {\bibinfo
  {journal} {Commun. Phys.}\ }\textbf {\bibinfo {volume} {108}},\ \bibinfo
  {pages} {233} (\bibinfo {year} {2022})}\BibitemShut {NoStop}%
\bibitem [{\citenamefont {Hagmann}\ and\ \citenamefont
  {Richards}(1995)}]{Hagmann1995}%
  \BibitemOpen
  \bibfield  {author} {\bibinfo {author} {\bibfnamefont {C.}~\bibnamefont
  {Hagmann}}\ and\ \bibinfo {author} {\bibfnamefont {P.~L.}\ \bibnamefont
  {Richards}},\ }\bibfield  {title} {\bibinfo {title} {Adiabatic
  demagnetization refrigerators for small laboratory experiments and space
  astronomy},\ }\href {https://doi.org/10.1016/0011-2275(95)95348-I} {\bibfield
   {journal} {\bibinfo  {journal} {Cryogenics}\ }\textbf {\bibinfo {volume}
  {35}},\ \bibinfo {pages} {303} (\bibinfo {year} {1995})}\BibitemShut
  {NoStop}%
\bibitem [{\citenamefont {Shirron}(2014)}]{Shirron2014}%
  \BibitemOpen
  \bibfield  {author} {\bibinfo {author} {\bibfnamefont {P.~J.}\ \bibnamefont
  {Shirron}},\ }\bibfield  {title} {\bibinfo {title} {Applications of the
  magnetocaloric effect in single-stage, multi-stage and continuous adiabatic
  demagnetization refrigerators},\ }\href
  {https://doi.org/https://doi.org/10.1016/j.cryogenics.2014.03.014} {\bibfield
   {journal} {\bibinfo  {journal} {Cryogenics}\ }\textbf {\bibinfo {volume}
  {62}},\ \bibinfo {pages} {130} (\bibinfo {year} {2014})}\BibitemShut
  {NoStop}%
\bibitem [{\citenamefont {Jahromi}\ \emph {et~al.}(2019)\citenamefont
  {Jahromi}, \citenamefont {Shirron},\ and\ \citenamefont
  {DiPirro}}]{Jahromi2019nasa}%
  \BibitemOpen
  \bibfield  {author} {\bibinfo {author} {\bibfnamefont {A.~E.}\ \bibnamefont
  {Jahromi}}, \bibinfo {author} {\bibfnamefont {P.~J.}\ \bibnamefont
  {Shirron}},\ and\ \bibinfo {author} {\bibfnamefont {M.~J.}\ \bibnamefont
  {DiPirro}},\ }\href@noop {} {\emph {\bibinfo {title} {{Sub-Kelvin Cooling
  Systems for Quantum Computers}}}},\ \bibinfo {type} {Tech. Rep.}\ (\bibinfo
  {institution} {NASA Goddard Space Flight Center Greenbelt, MD, United
  States},\ \bibinfo {year} {2019})\BibitemShut {NoStop}%
\bibitem [{\citenamefont {Schiffer}\ \emph {et~al.}(1994)\citenamefont
  {Schiffer}, \citenamefont {Ramirez}, \citenamefont {Huse},\ and\
  \citenamefont {Valentino}}]{Schiffer1994}%
  \BibitemOpen
  \bibfield  {author} {\bibinfo {author} {\bibfnamefont {P.}~\bibnamefont
  {Schiffer}}, \bibinfo {author} {\bibfnamefont {A.~P.}\ \bibnamefont
  {Ramirez}}, \bibinfo {author} {\bibfnamefont {D.~A.}\ \bibnamefont {Huse}},\
  and\ \bibinfo {author} {\bibfnamefont {A.~J.}\ \bibnamefont {Valentino}},\
  }\bibfield  {title} {\bibinfo {title} {Investigation of the field induced
  antiferromagnetic phase transition in the frustrated magnet: Gadolinium
  gallium garnet},\ }\href {https://doi.org/10.1103/PhysRevLett.73.2500}
  {\bibfield  {journal} {\bibinfo  {journal} {Phys. Rev. Lett.}\ }\textbf
  {\bibinfo {volume} {73}},\ \bibinfo {pages} {2500} (\bibinfo {year}
  {1994})}\BibitemShut {NoStop}%
\bibitem [{\citenamefont {Sanders}\ \emph {et~al.}(2017)\citenamefont
  {Sanders}, \citenamefont {Cevallos},\ and\ \citenamefont
  {Cava}}]{Sanders2017}%
  \BibitemOpen
  \bibfield  {author} {\bibinfo {author} {\bibfnamefont {M.~B.}\ \bibnamefont
  {Sanders}}, \bibinfo {author} {\bibfnamefont {F.~A.}\ \bibnamefont
  {Cevallos}},\ and\ \bibinfo {author} {\bibfnamefont {R.~J.}\ \bibnamefont
  {Cava}},\ }\bibfield  {title} {\bibinfo {title} {Magnetism in the
  {{KBaRE}({BO}$_3$)$_2$({RE}{\hspace{0.167em}}{\hspace{0.167em}}={\hspace{0.167em}}{\hspace{0.167em}}Sm,
  Eu, Gd, Tb, Dy, Ho, Er, Tm, Yb, Lu)} series: materials with a triangular rare
  earth lattice},\ }\href {https://doi.org/10.1088/2053-1591/aa60a2} {\bibfield
   {journal} {\bibinfo  {journal} {Mater. Res. Express}\ }\textbf {\bibinfo
  {volume} {4}},\ \bibinfo {pages} {036102} (\bibinfo {year}
  {2017})}\BibitemShut {NoStop}%
\bibitem [{\citenamefont {Guo}\ \emph {et~al.}(2019{\natexlab{a}})\citenamefont
  {Guo}, \citenamefont {Kong}, \citenamefont {Cevallos}, \citenamefont
  {Stolze},\ and\ \citenamefont {Cava}}]{Guo2019KBYB}%
  \BibitemOpen
  \bibfield  {author} {\bibinfo {author} {\bibfnamefont {S.}~\bibnamefont
  {Guo}}, \bibinfo {author} {\bibfnamefont {T.}~\bibnamefont {Kong}}, \bibinfo
  {author} {\bibfnamefont {F.~A.}\ \bibnamefont {Cevallos}}, \bibinfo {author}
  {\bibfnamefont {K.}~\bibnamefont {Stolze}},\ and\ \bibinfo {author}
  {\bibfnamefont {R.}~\bibnamefont {Cava}},\ }\bibfield  {title} {\bibinfo
  {title} {Crystal growth, crystal structure and anisotropic magnetic
  properties of {KBaR(BO$_3$)$_2$ (R=Y, Gd, Tb, Dy, Ho, Tm, Yb and Lu)}
  triangular lattice materials},\ }\href
  {https://doi.org/https://doi.org/10.1016/j.jmmm.2018.10.037} {\bibfield
  {journal} {\bibinfo  {journal} {J. Magn. Magn. Mater.}\ }\textbf {\bibinfo
  {volume} {472}},\ \bibinfo {pages} {104} (\bibinfo {year}
  {2019}{\natexlab{a}})}\BibitemShut {NoStop}%
\bibitem [{\citenamefont {Hasenfratz}\ and\ \citenamefont
  {Niedermayer}(1993)}]{Cv_Hasenfratz_1993}%
  \BibitemOpen
  \bibfield  {author} {\bibinfo {author} {\bibfnamefont {P.}~\bibnamefont
  {Hasenfratz}}\ and\ \bibinfo {author} {\bibfnamefont {F.}~\bibnamefont
  {Niedermayer}},\ }\bibfield  {title} {\bibinfo {title} {Finite size and
  temperature effects in the {AF} heisenberg model},\ }\href
  {https://doi.org/10.1007/bf01309171} {\bibfield  {journal} {\bibinfo
  {journal} {Z. Phys., B Condens. matter}\ }\textbf {\bibinfo {volume} {92}},\
  \bibinfo {pages} {91} (\bibinfo {year} {1993})}\BibitemShut {NoStop}%
\bibitem [{\citenamefont {Sandvik}\ and\ \citenamefont
  {Hamer}(1999)}]{Cv_Sandvik1999}%
  \BibitemOpen
  \bibfield  {author} {\bibinfo {author} {\bibfnamefont {A.~W.}\ \bibnamefont
  {Sandvik}}\ and\ \bibinfo {author} {\bibfnamefont {C.~J.}\ \bibnamefont
  {Hamer}},\ }\bibfield  {title} {\bibinfo {title} {Ground-state parameters,
  finite-size scaling, and low-temperature properties of the two-dimensional
  {$S=\frac{1}{2}$} $\mathrm{XY}$ model},\ }\href
  {https://doi.org/10.1103/PhysRevB.60.6588} {\bibfield  {journal} {\bibinfo
  {journal} {Phys. Rev. B}\ }\textbf {\bibinfo {volume} {60}},\ \bibinfo
  {pages} {6588} (\bibinfo {year} {1999})}\BibitemShut {NoStop}%
\bibitem [{\citenamefont {Paddison}\ \emph {et~al.}(2015)\citenamefont
  {Paddison}, \citenamefont {Jacobsen}, \citenamefont {Petrenko}, \citenamefont
  {Fern{\'a}ndez-D{\'\i}az}, \citenamefont {Deen},\ and\ \citenamefont
  {Goodwin}}]{paddison2015hidden}%
  \BibitemOpen
  \bibfield  {author} {\bibinfo {author} {\bibfnamefont {J.~A.}\ \bibnamefont
  {Paddison}}, \bibinfo {author} {\bibfnamefont {H.}~\bibnamefont {Jacobsen}},
  \bibinfo {author} {\bibfnamefont {O.~A.}\ \bibnamefont {Petrenko}}, \bibinfo
  {author} {\bibfnamefont {M.~T.}\ \bibnamefont {Fern{\'a}ndez-D{\'\i}az}},
  \bibinfo {author} {\bibfnamefont {P.~P.}\ \bibnamefont {Deen}},\ and\
  \bibinfo {author} {\bibfnamefont {A.~L.}\ \bibnamefont {Goodwin}},\
  }\bibfield  {title} {\bibinfo {title} {Hidden order in spin-liquid
  {Gd$_3$Ga$_5$O$_{12}$}},\ }\href {https://doi.org/10.1126/science.aaa5326}
  {\bibfield  {journal} {\bibinfo  {journal} {Science}\ }\textbf {\bibinfo
  {volume} {350}},\ \bibinfo {pages} {179} (\bibinfo {year}
  {2015})}\BibitemShut {NoStop}%
\bibitem [{\citenamefont {Zhu}\ \emph {et~al.}(2003)\citenamefont {Zhu},
  \citenamefont {Garst}, \citenamefont {Rosch},\ and\ \citenamefont
  {Si}}]{Zhu2003}%
  \BibitemOpen
  \bibfield  {author} {\bibinfo {author} {\bibfnamefont {L.~J.}\ \bibnamefont
  {Zhu}}, \bibinfo {author} {\bibfnamefont {M.}~\bibnamefont {Garst}}, \bibinfo
  {author} {\bibfnamefont {A.}~\bibnamefont {Rosch}},\ and\ \bibinfo {author}
  {\bibfnamefont {Q.~M.}\ \bibnamefont {Si}},\ }\bibfield  {title} {\bibinfo
  {title} {{Universally Diverging Gr{\"u}neisen Parameter and the
  Magnetocaloric Effect Close to Quantum Critical Points}},\ }\href
  {https://doi.org/10.1103/PhysRevLett.91.066404} {\bibfield  {journal}
  {\bibinfo  {journal} {Phys. Rev. Lett.}\ }\textbf {\bibinfo {volume} {91}},\
  \bibinfo {pages} {066404} (\bibinfo {year} {2003})}\BibitemShut {NoStop}%
\bibitem [{\citenamefont {Xiang}\ \emph {et~al.}(2017)\citenamefont {Xiang},
  \citenamefont {Chen}, \citenamefont {Li}, \citenamefont {Sheng},
  \citenamefont {Su}, \citenamefont {Cheng}, \citenamefont {Chen},\ and\
  \citenamefont {Chen}}]{Xiang2017}%
  \BibitemOpen
  \bibfield  {author} {\bibinfo {author} {\bibfnamefont {J.-S.}\ \bibnamefont
  {Xiang}}, \bibinfo {author} {\bibfnamefont {C.}~\bibnamefont {Chen}},
  \bibinfo {author} {\bibfnamefont {W.}~\bibnamefont {Li}}, \bibinfo {author}
  {\bibfnamefont {X.-L.}\ \bibnamefont {Sheng}}, \bibinfo {author}
  {\bibfnamefont {N.}~\bibnamefont {Su}}, \bibinfo {author} {\bibfnamefont
  {Z.-H.}\ \bibnamefont {Cheng}}, \bibinfo {author} {\bibfnamefont
  {Q.}~\bibnamefont {Chen}},\ and\ \bibinfo {author} {\bibfnamefont {Z.-Y.}\
  \bibnamefont {Chen}},\ }\bibfield  {title} {\bibinfo {title}
  {Criticality-enhanced magnetocaloric effect in quantum spin chain material
  copper nitrate},\ }\href {https://doi.org/10.1038/srep44643} {\bibfield
  {journal} {\bibinfo  {journal} {Sci. Rep.}\ }\textbf {\bibinfo {volume}
  {7}},\ \bibinfo {pages} {44643} (\bibinfo {year} {2017})}\BibitemShut
  {NoStop}%
\bibitem [{\citenamefont {Garst}\ and\ \citenamefont
  {Rosch}(2005)}]{Garst2005}%
  \BibitemOpen
  \bibfield  {author} {\bibinfo {author} {\bibfnamefont {M.}~\bibnamefont
  {Garst}}\ and\ \bibinfo {author} {\bibfnamefont {A.}~\bibnamefont {Rosch}},\
  }\bibfield  {title} {\bibinfo {title} {Sign change of the {Gr{\"u}neisen}
  parameter and magnetocaloric effect near quantum critical points},\ }\href
  {https://doi.org/10.1103/PhysRevB.72.205129} {\bibfield  {journal} {\bibinfo
  {journal} {Phys. Rev. B}\ }\textbf {\bibinfo {volume} {72}},\ \bibinfo
  {pages} {205129} (\bibinfo {year} {2005})}\BibitemShut {NoStop}%
\bibitem [{\citenamefont {Liu}\ \emph {et~al.}(2021)\citenamefont {Liu},
  \citenamefont {Liu}, \citenamefont {Gao}, \citenamefont {Jin}, \citenamefont
  {He}, \citenamefont {Sheng}, \citenamefont {Jin}, \citenamefont {Chen},\ and\
  \citenamefont {Li}}]{Liu2021}%
  \BibitemOpen
  \bibfield  {author} {\bibinfo {author} {\bibfnamefont {T.}~\bibnamefont
  {Liu}}, \bibinfo {author} {\bibfnamefont {X.-Y.}\ \bibnamefont {Liu}},
  \bibinfo {author} {\bibfnamefont {Y.}~\bibnamefont {Gao}}, \bibinfo {author}
  {\bibfnamefont {H.}~\bibnamefont {Jin}}, \bibinfo {author} {\bibfnamefont
  {J.}~\bibnamefont {He}}, \bibinfo {author} {\bibfnamefont {X.-L.}\
  \bibnamefont {Sheng}}, \bibinfo {author} {\bibfnamefont {W.}~\bibnamefont
  {Jin}}, \bibinfo {author} {\bibfnamefont {Z.}~\bibnamefont {Chen}},\ and\
  \bibinfo {author} {\bibfnamefont {W.}~\bibnamefont {Li}},\ }\bibfield
  {title} {\bibinfo {title} {Significant inverse magnetocaloric effect induced
  by quantum criticality},\ }\href
  {https://doi.org/10.1103/PhysRevResearch.3.033094} {\bibfield  {journal}
  {\bibinfo  {journal} {Phys. Rev. Research}\ }\textbf {\bibinfo {volume}
  {3}},\ \bibinfo {pages} {033094} (\bibinfo {year} {2021})}\BibitemShut
  {NoStop}%
\bibitem [{\citenamefont {Tokiwa}\ \emph {et~al.}(2009)\citenamefont {Tokiwa},
  \citenamefont {Radu}, \citenamefont {Geibel}, \citenamefont {Steglich},\ and\
  \citenamefont {Gegenwart}}]{Tokiwa2009}%
  \BibitemOpen
  \bibfield  {author} {\bibinfo {author} {\bibfnamefont {Y.}~\bibnamefont
  {Tokiwa}}, \bibinfo {author} {\bibfnamefont {T.}~\bibnamefont {Radu}},
  \bibinfo {author} {\bibfnamefont {C.}~\bibnamefont {Geibel}}, \bibinfo
  {author} {\bibfnamefont {F.}~\bibnamefont {Steglich}},\ and\ \bibinfo
  {author} {\bibfnamefont {P.}~\bibnamefont {Gegenwart}},\ }\bibfield  {title}
  {\bibinfo {title} {{Divergence of the Magnetic Gr\"uneisen Ratio at the
  Field-Induced Quantum Critical Point in {YbRh}$_{2}${Si}$_{2}$}},\ }\href
  {https://doi.org/10.1103/PhysRevLett.102.066401} {\bibfield  {journal}
  {\bibinfo  {journal} {Phys. Rev. Lett.}\ }\textbf {\bibinfo {volume} {102}},\
  \bibinfo {pages} {066401} (\bibinfo {year} {2009})}\BibitemShut {NoStop}%
\bibitem [{\citenamefont {Jang}\ \emph {et~al.}(2015)\citenamefont {Jang},
  \citenamefont {Gruner}, \citenamefont {Steppke}, \citenamefont {Mitsumoto},
  \citenamefont {Geibel},\ and\ \citenamefont {Brando}}]{Jang2015}%
  \BibitemOpen
  \bibfield  {author} {\bibinfo {author} {\bibfnamefont {D.}~\bibnamefont
  {Jang}}, \bibinfo {author} {\bibfnamefont {T.}~\bibnamefont {Gruner}},
  \bibinfo {author} {\bibfnamefont {A.}~\bibnamefont {Steppke}}, \bibinfo
  {author} {\bibfnamefont {K.}~\bibnamefont {Mitsumoto}}, \bibinfo {author}
  {\bibfnamefont {C.}~\bibnamefont {Geibel}},\ and\ \bibinfo {author}
  {\bibfnamefont {M.}~\bibnamefont {Brando}},\ }\bibfield  {title} {\bibinfo
  {title} {Large magnetocaloric effect and adiabatic demagnetization
  refrigeration with {YbPt$_2$Sn}},\ }\href
  {https://doi.org/10.1038/ncomms9680} {\bibfield  {journal} {\bibinfo
  {journal} {Nat. Commun.}\ }\textbf {\bibinfo {volume} {6}},\ \bibinfo {pages}
  {8680} (\bibinfo {year} {2015})}\BibitemShut {NoStop}%
\bibitem [{\citenamefont {{Tokiwa}}\ \emph {et~al.}(2016)\citenamefont
  {{Tokiwa}}, \citenamefont {{Piening}}, \citenamefont {{Jeevan}},
  \citenamefont {{Budko}}, \citenamefont {{Canfield}},\ and\ \citenamefont
  {{Gegenwart}}}]{Tokiwa2016}%
  \BibitemOpen
  \bibfield  {author} {\bibinfo {author} {\bibfnamefont {Y.}~\bibnamefont
  {{Tokiwa}}}, \bibinfo {author} {\bibfnamefont {B.}~\bibnamefont {{Piening}}},
  \bibinfo {author} {\bibfnamefont {H.~S.}\ \bibnamefont {{Jeevan}}}, \bibinfo
  {author} {\bibfnamefont {S.~L.}\ \bibnamefont {{Budko}}}, \bibinfo {author}
  {\bibfnamefont {P.~C.}\ \bibnamefont {{Canfield}}},\ and\ \bibinfo {author}
  {\bibfnamefont {P.}~\bibnamefont {{Gegenwart}}},\ }\bibfield  {title}
  {\bibinfo {title} {{Super-heavy electron material as metallic refrigerant for
  adiabatic demagnetization cooling}},\ }\href
  {https://doi.org/10.1126/sciadv.1600835} {\bibfield  {journal} {\bibinfo
  {journal} {Sci. Adv.}\ }\textbf {\bibinfo {volume} {2}},\ \bibinfo {pages}
  {e1600835} (\bibinfo {year} {2016})}\BibitemShut {NoStop}%
\bibitem [{\citenamefont {Gegenwart}(2016)}]{Gegenwart2016}%
  \BibitemOpen
  \bibfield  {author} {\bibinfo {author} {\bibfnamefont {P.}~\bibnamefont
  {Gegenwart}},\ }\bibfield  {title} {\bibinfo {title} {Gr{\"u}neisen parameter
  studies on heavy fermion quantum criticality},\ }\href
  {https://doi.org/10.1088/0034-4885/79/11/114502} {\bibfield  {journal}
  {\bibinfo  {journal} {Rep. Prog. Phys.}\ }\textbf {\bibinfo {volume} {79}},\
  \bibinfo {pages} {114502} (\bibinfo {year} {2016})}\BibitemShut {NoStop}%
\bibitem [{\citenamefont {Shimura}\ \emph {et~al.}(2022)\citenamefont
  {Shimura}, \citenamefont {Watanabe}, \citenamefont {Taniguchi}, \citenamefont
  {Osato}, \citenamefont {Yamamoto}, \citenamefont {Kusanose}, \citenamefont
  {Umeo}, \citenamefont {Fujita}, \citenamefont {Onimaru},\ and\ \citenamefont
  {Takabatake}}]{shimura2022magnetic}%
  \BibitemOpen
  \bibfield  {author} {\bibinfo {author} {\bibfnamefont {Y.}~\bibnamefont
  {Shimura}}, \bibinfo {author} {\bibfnamefont {K.}~\bibnamefont {Watanabe}},
  \bibinfo {author} {\bibfnamefont {T.}~\bibnamefont {Taniguchi}}, \bibinfo
  {author} {\bibfnamefont {K.}~\bibnamefont {Osato}}, \bibinfo {author}
  {\bibfnamefont {R.}~\bibnamefont {Yamamoto}}, \bibinfo {author}
  {\bibfnamefont {Y.}~\bibnamefont {Kusanose}}, \bibinfo {author}
  {\bibfnamefont {K.}~\bibnamefont {Umeo}}, \bibinfo {author} {\bibfnamefont
  {M.}~\bibnamefont {Fujita}}, \bibinfo {author} {\bibfnamefont
  {T.}~\bibnamefont {Onimaru}},\ and\ \bibinfo {author} {\bibfnamefont
  {T.}~\bibnamefont {Takabatake}},\ }\bibfield  {title} {\bibinfo {title}
  {Magnetic refrigeration down to {0.2 K} by heavy fermion metal
  {YbCu$_4$Ni}},\ }\href {https://doi.org/10.1063/5.0064355} {\bibfield
  {journal} {\bibinfo  {journal} {J. of Appl. Phys.}\ }\textbf {\bibinfo
  {volume} {131}},\ \bibinfo {pages} {013903} (\bibinfo {year}
  {2022})}\BibitemShut {NoStop}%
\bibitem [{\citenamefont {Honecker}\ and\ \citenamefont
  {Wessel}(2009)}]{Honecker2009}%
  \BibitemOpen
  \bibfield  {author} {\bibinfo {author} {\bibfnamefont {A.}~\bibnamefont
  {Honecker}}\ and\ \bibinfo {author} {\bibfnamefont {S.}~\bibnamefont
  {Wessel}},\ }\bibfield  {title} {\bibinfo {title} {Magnetocaloric effect in
  quantum spin-s chains},\ }\href {http://dx.doi.org/10.5488/CMP.12.3.399}
  {\bibfield  {journal} {\bibinfo  {journal} {Condens. Matter Phys.}\ }\textbf
  {\bibinfo {volume} {12}},\ \bibinfo {pages} {399} (\bibinfo {year}
  {2009})}\BibitemShut {NoStop}%
\bibitem [{\citenamefont {Wolf}\ \emph {et~al.}(2011)\citenamefont {Wolf},
  \citenamefont {Tsui}, \citenamefont {Jaiswal-Nagar}, \citenamefont {Tutsch},
  \citenamefont {Honecker}, \citenamefont {Removi{\'c}-Langer}, \citenamefont
  {Hofmann}, \citenamefont {Prokofiev}, \citenamefont {Assmus}, \citenamefont
  {Donath},\ and\ \citenamefont {Lang}}]{Wolf2011}%
  \BibitemOpen
  \bibfield  {author} {\bibinfo {author} {\bibfnamefont {B.}~\bibnamefont
  {Wolf}}, \bibinfo {author} {\bibfnamefont {Y.}~\bibnamefont {Tsui}}, \bibinfo
  {author} {\bibfnamefont {D.}~\bibnamefont {Jaiswal-Nagar}}, \bibinfo {author}
  {\bibfnamefont {U.}~\bibnamefont {Tutsch}}, \bibinfo {author} {\bibfnamefont
  {A.}~\bibnamefont {Honecker}}, \bibinfo {author} {\bibfnamefont
  {K.}~\bibnamefont {Removi{\'c}-Langer}}, \bibinfo {author} {\bibfnamefont
  {G.}~\bibnamefont {Hofmann}}, \bibinfo {author} {\bibfnamefont
  {A.}~\bibnamefont {Prokofiev}}, \bibinfo {author} {\bibfnamefont
  {W.}~\bibnamefont {Assmus}}, \bibinfo {author} {\bibfnamefont
  {G.}~\bibnamefont {Donath}},\ and\ \bibinfo {author} {\bibfnamefont
  {M.}~\bibnamefont {Lang}},\ }\bibfield  {title} {\bibinfo {title}
  {Magnetocaloric effect and magnetic cooling near a field-induced
  quantum-critical point},\ }\href {https://doi.org/10.1073/pnas.1017047108}
  {\bibfield  {journal} {\bibinfo  {journal} {Proc. Natl. Acad. Sci.}\ }\textbf
  {\bibinfo {volume} {108}},\ \bibinfo {pages} {6862} (\bibinfo {year}
  {2011})}\BibitemShut {NoStop}%
\bibitem [{\citenamefont {Lang}\ \emph {et~al.}(2012)\citenamefont {Lang},
  \citenamefont {Wolf}, \citenamefont {Honecker}, \citenamefont {Tsui},
  \citenamefont {Jaiswal-Nagar}, \citenamefont {Tutsch}, \citenamefont
  {Hofmann}, \citenamefont {Prokofiev}, \citenamefont {Cong}, \citenamefont
  {Kr{\"u}ger}, \citenamefont {Ritter},\ and\ \citenamefont
  {Assmus}}]{Lang2012}%
  \BibitemOpen
  \bibfield  {author} {\bibinfo {author} {\bibfnamefont {M.}~\bibnamefont
  {Lang}}, \bibinfo {author} {\bibfnamefont {B.}~\bibnamefont {Wolf}}, \bibinfo
  {author} {\bibfnamefont {A.}~\bibnamefont {Honecker}}, \bibinfo {author}
  {\bibfnamefont {Y.}~\bibnamefont {Tsui}}, \bibinfo {author} {\bibfnamefont
  {D.}~\bibnamefont {Jaiswal-Nagar}}, \bibinfo {author} {\bibfnamefont
  {U.}~\bibnamefont {Tutsch}}, \bibinfo {author} {\bibfnamefont
  {G.}~\bibnamefont {Hofmann}}, \bibinfo {author} {\bibfnamefont
  {A.}~\bibnamefont {Prokofiev}}, \bibinfo {author} {\bibfnamefont {P.~T.}\
  \bibnamefont {Cong}}, \bibinfo {author} {\bibfnamefont {N.}~\bibnamefont
  {Kr{\"u}ger}}, \bibinfo {author} {\bibfnamefont {F.}~\bibnamefont {Ritter}},\
  and\ \bibinfo {author} {\bibfnamefont {W.}~\bibnamefont {Assmus}},\
  }\bibfield  {title} {\bibinfo {title} {Magnetic cooling through quantum
  criticality},\ }\href {https://doi.org/10.1088/1742-6596/400/3/032043}
  {\bibfield  {journal} {\bibinfo  {journal} {J. Phys.: Conf. Series}\ }\textbf
  {\bibinfo {volume} {400}},\ \bibinfo {pages} {032043} (\bibinfo {year}
  {2012})}\BibitemShut {NoStop}%
\bibitem [{\citenamefont {Bachus}\ \emph {et~al.}(2020)\citenamefont {Bachus},
  \citenamefont {Kaib}, \citenamefont {Tokiwa}, \citenamefont {Jesche},
  \citenamefont {Tsurkan}, \citenamefont {Loidl}, \citenamefont {Winter},
  \citenamefont {Tsirlin}, \citenamefont {Valent\'{\i}},\ and\ \citenamefont
  {Gegenwart}}]{Bachus2020}%
  \BibitemOpen
  \bibfield  {author} {\bibinfo {author} {\bibfnamefont {S.}~\bibnamefont
  {Bachus}}, \bibinfo {author} {\bibfnamefont {D.~A.~S.}\ \bibnamefont {Kaib}},
  \bibinfo {author} {\bibfnamefont {Y.}~\bibnamefont {Tokiwa}}, \bibinfo
  {author} {\bibfnamefont {A.}~\bibnamefont {Jesche}}, \bibinfo {author}
  {\bibfnamefont {V.}~\bibnamefont {Tsurkan}}, \bibinfo {author} {\bibfnamefont
  {A.}~\bibnamefont {Loidl}}, \bibinfo {author} {\bibfnamefont {S.~M.}\
  \bibnamefont {Winter}}, \bibinfo {author} {\bibfnamefont {A.~A.}\
  \bibnamefont {Tsirlin}}, \bibinfo {author} {\bibfnamefont {R.}~\bibnamefont
  {Valent\'{\i}}},\ and\ \bibinfo {author} {\bibfnamefont {P.}~\bibnamefont
  {Gegenwart}},\ }\bibfield  {title} {\bibinfo {title} {Thermodynamic
  perspective on field-induced behavior of {$\alpha$ RuCl$_3$}},\ }\href
  {https://doi.org/10.1103/PhysRevLett.125.097203} {\bibfield  {journal}
  {\bibinfo  {journal} {Phys. Rev. Lett.}\ }\textbf {\bibinfo {volume} {125}},\
  \bibinfo {pages} {097203} (\bibinfo {year} {2020})}\BibitemShut {NoStop}%
\bibitem [{\citenamefont {Jos\'e}\ \emph {et~al.}(1977)\citenamefont {Jos\'e},
  \citenamefont {Kadanoff}, \citenamefont {Kirkpatrick},\ and\ \citenamefont
  {Nelson}}]{Jose1977}%
  \BibitemOpen
  \bibfield  {author} {\bibinfo {author} {\bibfnamefont {J.~V.}\ \bibnamefont
  {Jos\'e}}, \bibinfo {author} {\bibfnamefont {L.~P.}\ \bibnamefont
  {Kadanoff}}, \bibinfo {author} {\bibfnamefont {S.}~\bibnamefont
  {Kirkpatrick}},\ and\ \bibinfo {author} {\bibfnamefont {D.~R.}\ \bibnamefont
  {Nelson}},\ }\bibfield  {title} {\bibinfo {title} {Renormalization, vortices,
  and symmetry-breaking perturbations in the two-dimensional planar model},\
  }\href {https://doi.org/10.1103/PhysRevB.16.1217} {\bibfield  {journal}
  {\bibinfo  {journal} {Phys. Rev. B}\ }\textbf {\bibinfo {volume} {16}},\
  \bibinfo {pages} {1217} (\bibinfo {year} {1977})}\BibitemShut {NoStop}%
\bibitem [{\citenamefont {Moessner}\ \emph {et~al.}(2000)\citenamefont
  {Moessner}, \citenamefont {Sondhi},\ and\ \citenamefont
  {Chandra}}]{Moessner2000}%
  \BibitemOpen
  \bibfield  {author} {\bibinfo {author} {\bibfnamefont {R.}~\bibnamefont
  {Moessner}}, \bibinfo {author} {\bibfnamefont {S.~L.}\ \bibnamefont
  {Sondhi}},\ and\ \bibinfo {author} {\bibfnamefont {P.}~\bibnamefont
  {Chandra}},\ }\bibfield  {title} {\bibinfo {title} {Two-dimensional periodic
  frustrated ising models in a transverse field},\ }\href
  {https://doi.org/10.1103/PhysRevLett.84.4457} {\bibfield  {journal} {\bibinfo
   {journal} {Phys. Rev. Lett.}\ }\textbf {\bibinfo {volume} {84}},\ \bibinfo
  {pages} {4457} (\bibinfo {year} {2000})}\BibitemShut {NoStop}%
\bibitem [{\citenamefont {Moessner}\ and\ \citenamefont
  {Sondhi}(2001)}]{Moessner2001}%
  \BibitemOpen
  \bibfield  {author} {\bibinfo {author} {\bibfnamefont {R.}~\bibnamefont
  {Moessner}}\ and\ \bibinfo {author} {\bibfnamefont {S.~L.}\ \bibnamefont
  {Sondhi}},\ }\bibfield  {title} {\bibinfo {title} {Ising models of quantum
  frustration},\ }\href {https://doi.org/10.1103/PhysRevB.63.224401} {\bibfield
   {journal} {\bibinfo  {journal} {Phys. Rev. B}\ }\textbf {\bibinfo {volume}
  {63}},\ \bibinfo {pages} {224401} (\bibinfo {year} {2001})}\BibitemShut
  {NoStop}%
\bibitem [{\citenamefont {Isakov}\ and\ \citenamefont
  {Moessner}(2003)}]{Isakov2003}%
  \BibitemOpen
  \bibfield  {author} {\bibinfo {author} {\bibfnamefont {S.~V.}\ \bibnamefont
  {Isakov}}\ and\ \bibinfo {author} {\bibfnamefont {R.}~\bibnamefont
  {Moessner}},\ }\bibfield  {title} {\bibinfo {title} {Interplay of quantum and
  thermal fluctuations in a frustrated magnet},\ }\href
  {https://doi.org/10.1103/PhysRevB.68.104409} {\bibfield  {journal} {\bibinfo
  {journal} {Phys. Rev. B}\ }\textbf {\bibinfo {volume} {68}},\ \bibinfo
  {pages} {104409} (\bibinfo {year} {2003})}\BibitemShut {NoStop}%
\bibitem [{\citenamefont {Baek}\ \emph {et~al.}(2011)\citenamefont {Baek},
  \citenamefont {Minnhagen},\ and\ \citenamefont {Kim}}]{KTDP_Baek2011PRB}%
  \BibitemOpen
  \bibfield  {author} {\bibinfo {author} {\bibfnamefont {S.~K.}\ \bibnamefont
  {Baek}}, \bibinfo {author} {\bibfnamefont {P.}~\bibnamefont {Minnhagen}},\
  and\ \bibinfo {author} {\bibfnamefont {B.~J.}\ \bibnamefont {Kim}},\
  }\bibfield  {title} {\bibinfo {title} {{Kosterlitz-Thouless} transition of
  magnetic dipoles on the two-dimensional plane},\ }\href
  {https://doi.org/10.1103/PhysRevB.83.184409} {\bibfield  {journal} {\bibinfo
  {journal} {Phys. Rev. B}\ }\textbf {\bibinfo {volume} {83}},\ \bibinfo
  {pages} {184409} (\bibinfo {year} {2011})}\BibitemShut {NoStop}%
\bibitem [{\citenamefont {Gottlob}\ and\ \citenamefont
  {Hasenbusch}(1993)}]{Gottlob1993}%
  \BibitemOpen
  \bibfield  {author} {\bibinfo {author} {\bibfnamefont {A.~P.}\ \bibnamefont
  {Gottlob}}\ and\ \bibinfo {author} {\bibfnamefont {M.}~\bibnamefont
  {Hasenbusch}},\ }\bibfield  {title} {\bibinfo {title} {Critical behaviour of
  the {3D XY-model: a Monte Carlo} study},\ }\href
  {https://doi.org/https://doi.org/10.1016/0378-4371(93)90131-M} {\bibfield
  {journal} {\bibinfo  {journal} {Physica A: Statistical Mechanics and its
  Applications}\ }\textbf {\bibinfo {volume} {201}},\ \bibinfo {pages} {593}
  (\bibinfo {year} {1993})}\BibitemShut {NoStop}%
\bibitem [{\citenamefont {Numazawa}\ \emph {et~al.}(2006)\citenamefont
  {Numazawa}, \citenamefont {Kamiya}, \citenamefont {Shirron}, \citenamefont
  {DiPirro},\ and\ \citenamefont {Matsumoto}}]{GLF2006}%
  \BibitemOpen
  \bibfield  {author} {\bibinfo {author} {\bibfnamefont {T.}~\bibnamefont
  {Numazawa}}, \bibinfo {author} {\bibfnamefont {K.}~\bibnamefont {Kamiya}},
  \bibinfo {author} {\bibfnamefont {P.}~\bibnamefont {Shirron}}, \bibinfo
  {author} {\bibfnamefont {M.}~\bibnamefont {DiPirro}},\ and\ \bibinfo {author}
  {\bibfnamefont {K.}~\bibnamefont {Matsumoto}},\ }\bibfield  {title} {\bibinfo
  {title} {Magnetocaloric effect of polycrystal {GdLiF$_4$} for adiabatic
  magnetic refrigeration},\ }\href {https://doi.org/10.1063/1.2355309}
  {\bibfield  {journal} {\bibinfo  {journal} {AIP Conference Proceedings}\
  }\textbf {\bibinfo {volume} {850}},\ \bibinfo {pages} {1579} (\bibinfo {year}
  {2006})}\BibitemShut {NoStop}%
\bibitem [{\citenamefont {Wikus}\ \emph {et~al.}(2014)\citenamefont {Wikus},
  \citenamefont {Canavan}, \citenamefont {Heine}, \citenamefont {Matsumoto},\
  and\ \citenamefont {Numazawa}}]{ADRdesign2014}%
  \BibitemOpen
  \bibfield  {author} {\bibinfo {author} {\bibfnamefont {P.}~\bibnamefont
  {Wikus}}, \bibinfo {author} {\bibfnamefont {E.}~\bibnamefont {Canavan}},
  \bibinfo {author} {\bibfnamefont {S.~T.}\ \bibnamefont {Heine}}, \bibinfo
  {author} {\bibfnamefont {K.}~\bibnamefont {Matsumoto}},\ and\ \bibinfo
  {author} {\bibfnamefont {T.}~\bibnamefont {Numazawa}},\ }\bibfield  {title}
  {\bibinfo {title} {Magnetocaloric materials and the optimization of cooling
  power density},\ }\href
  {https://doi.org/https://doi.org/10.1016/j.cryogenics.2014.04.005} {\bibfield
   {journal} {\bibinfo  {journal} {Cryogenics}\ }\textbf {\bibinfo {volume}
  {62}},\ \bibinfo {pages} {150} (\bibinfo {year} {2014})}\BibitemShut
  {NoStop}%
\bibitem [{\citenamefont {Vasiliev}\ \emph {et~al.}(2014)\citenamefont
  {Vasiliev}, \citenamefont {Tarkhov}, \citenamefont {Menshikov}, \citenamefont
  {Fedichev},\ and\ \citenamefont {Fischer}}]{KTDP_Vasiliev2014NJP}%
  \BibitemOpen
  \bibfield  {author} {\bibinfo {author} {\bibfnamefont {A.~Y.}\ \bibnamefont
  {Vasiliev}}, \bibinfo {author} {\bibfnamefont {A.~E.}\ \bibnamefont
  {Tarkhov}}, \bibinfo {author} {\bibfnamefont {L.~I.}\ \bibnamefont
  {Menshikov}}, \bibinfo {author} {\bibfnamefont {P.~O.}\ \bibnamefont
  {Fedichev}},\ and\ \bibinfo {author} {\bibfnamefont {U.~R.}\ \bibnamefont
  {Fischer}},\ }\bibfield  {title} {\bibinfo {title} {Universality of the
  {Berezinskii–Kosterlitz–Thouless} type of phase transition in the dipolar
  {XY}-model},\ }\href {https://doi.org/10.1088/1367-2630/16/5/053011}
  {\bibfield  {journal} {\bibinfo  {journal} {New J. Phys.}\ }\textbf {\bibinfo
  {volume} {16}},\ \bibinfo {pages} {053011} (\bibinfo {year}
  {2014})}\BibitemShut {NoStop}%
\bibitem [{\citenamefont {Guo}\ \emph {et~al.}(2019{\natexlab{b}})\citenamefont
  {Guo}, \citenamefont {Ghasemi}, \citenamefont {Broholm},\ and\ \citenamefont
  {Cava}}]{Guo2019NBYB}%
  \BibitemOpen
  \bibfield  {author} {\bibinfo {author} {\bibfnamefont {S.}~\bibnamefont
  {Guo}}, \bibinfo {author} {\bibfnamefont {A.}~\bibnamefont {Ghasemi}},
  \bibinfo {author} {\bibfnamefont {C.~L.}\ \bibnamefont {Broholm}},\ and\
  \bibinfo {author} {\bibfnamefont {R.~J.}\ \bibnamefont {Cava}},\ }\bibfield
  {title} {\bibinfo {title} {{Magnetism on ideal triangular lattices in
  NaBaYb(BO$_2$)$_2$}},\ }\href
  {https://doi.org/10.1103/PhysRevMaterials.3.094404} {\bibfield  {journal}
  {\bibinfo  {journal} {Phys. Rev. Mater.}\ }\textbf {\bibinfo {volume} {3}},\
  \bibinfo {pages} {094404} (\bibinfo {year} {2019}{\natexlab{b}})}\BibitemShut
  {NoStop}%
\bibitem [{\citenamefont {Tokiwa}\ \emph {et~al.}(2021)\citenamefont {Tokiwa},
  \citenamefont {Bachus}, \citenamefont {Kavita}, \citenamefont {Jesche},
  \citenamefont {Tsirlin},\ and\ \citenamefont {Gegenwart}}]{Tokiwa2021}%
  \BibitemOpen
  \bibfield  {author} {\bibinfo {author} {\bibfnamefont {Y.}~\bibnamefont
  {Tokiwa}}, \bibinfo {author} {\bibfnamefont {S.}~\bibnamefont {Bachus}},
  \bibinfo {author} {\bibfnamefont {K.}~\bibnamefont {Kavita}}, \bibinfo
  {author} {\bibfnamefont {A.}~\bibnamefont {Jesche}}, \bibinfo {author}
  {\bibfnamefont {A.~A.}\ \bibnamefont {Tsirlin}},\ and\ \bibinfo {author}
  {\bibfnamefont {P.}~\bibnamefont {Gegenwart}},\ }\bibfield  {title} {\bibinfo
  {title} {Frustrated magnet for adiabatic demagnetization cooling to
  milli-kelvin temperatures},\ }\href
  {https://doi.org/10.1038/s43246-021-00142-1} {\bibfield  {journal} {\bibinfo
  {journal} {Commun. Mater.}\ }\textbf {\bibinfo {volume} {2}},\ \bibinfo
  {pages} {42} (\bibinfo {year} {2021})}\BibitemShut {NoStop}%
\bibitem [{\citenamefont {{Jesche}}\ \emph {et~al.}()\citenamefont {{Jesche}},
  \citenamefont {{Winterhalter-Stocker}}, \citenamefont {{Hirschberger}},
  \citenamefont {{Bellon}}, \citenamefont {{Bachus}}, \citenamefont {{Tokiwa}},
  \citenamefont {{Tsirlin}},\ and\ \citenamefont
  {{Gegenwart}}}]{Jesche2022arXiv}%
  \BibitemOpen
  \bibfield  {author} {\bibinfo {author} {\bibfnamefont {A.}~\bibnamefont
  {{Jesche}}}, \bibinfo {author} {\bibfnamefont {N.}~\bibnamefont
  {{Winterhalter-Stocker}}}, \bibinfo {author} {\bibfnamefont {F.}~\bibnamefont
  {{Hirschberger}}}, \bibinfo {author} {\bibfnamefont {A.}~\bibnamefont
  {{Bellon}}}, \bibinfo {author} {\bibfnamefont {S.}~\bibnamefont {{Bachus}}},
  \bibinfo {author} {\bibfnamefont {Y.}~\bibnamefont {{Tokiwa}}}, \bibinfo
  {author} {\bibfnamefont {A.~A.}\ \bibnamefont {{Tsirlin}}},\ and\ \bibinfo
  {author} {\bibfnamefont {P.}~\bibnamefont {{Gegenwart}}},\ }\bibfield
  {title} {\bibinfo {title} {{Adiabatic demagnetization cooling well below the
  magnetic ordering temperature in the triangular antiferromagnet
  {KBaGd(BO$_3$)$_2$}}},\ }\href@noop {} {\ }\Eprint
  {https://arxiv.org/abs/2212.12483 (2022)} {arXiv:2212.12483 (2022)}
  \BibitemShut {NoStop}%
\bibitem [{\citenamefont {Hagmann}\ and\ \citenamefont
  {Richards}(1994)}]{Hagmann1994}%
  \BibitemOpen
  \bibfield  {author} {\bibinfo {author} {\bibfnamefont {C.}~\bibnamefont
  {Hagmann}}\ and\ \bibinfo {author} {\bibfnamefont {P.}~\bibnamefont
  {Richards}},\ }\bibfield  {title} {\bibinfo {title} {Two-stage magnetic
  refrigerator for astronomical applications with reservoir temperatures above
  4 {K}},\ }\href
  {https://doi.org/https://doi.org/10.1016/0011-2275(94)90172-4} {\bibfield
  {journal} {\bibinfo  {journal} {Cryogenics}\ }\textbf {\bibinfo {volume}
  {34}},\ \bibinfo {pages} {221} (\bibinfo {year} {1994})}\BibitemShut
  {NoStop}%
\bibitem [{\citenamefont {Sandvik}(2010)}]{Sandvik2010}%
  \BibitemOpen
  \bibfield  {author} {\bibinfo {author} {\bibfnamefont {A.~W.}\ \bibnamefont
  {Sandvik}},\ }\bibfield  {title} {\bibinfo {title} {Computational studies of
  quantum spin systems},\ }\href {https://doi.org/10.1063/1.3518900} {\bibfield
   {journal} {\bibinfo  {journal} {AIP Conf. Proc.}\ }\textbf {\bibinfo
  {volume} {1297}},\ \bibinfo {pages} {135} (\bibinfo {year}
  {2010})}\BibitemShut {NoStop}%
\bibitem [{\citenamefont {Creutz}(1987)}]{Overrelaxation}%
  \BibitemOpen
  \bibfield  {author} {\bibinfo {author} {\bibfnamefont {M.}~\bibnamefont
  {Creutz}},\ }\bibfield  {title} {\bibinfo {title} {Overrelaxation and monte
  carlo simulation},\ }\href {https://doi.org/10.1103/PhysRevD.36.515}
  {\bibfield  {journal} {\bibinfo  {journal} {Phys. Rev. D}\ }\textbf {\bibinfo
  {volume} {36}},\ \bibinfo {pages} {515} (\bibinfo {year} {1987})}\BibitemShut
  {NoStop}%
\end{thebibliography}%

% =========== Supplementary Materials =============
\clearpage
\linespread{1.2}
\setlength{\baselineskip}{15pt}

\newpage
\clearpage
\onecolumngrid
\mbox{}

%======================
\begin{center}
	{\large Supplementary Materials}
	$\,$\\ \textbf{\large{Dipolar Spin Liquid Ending with
			Quantum Critical Point in a Gd-based Triangular Magnet}}	
	$\,$\\
	Xiang \textit{et al.}
\end{center}
%=====================

\date{\today}

\setcounter{section}{0}
\setcounter{figure}{0}
\setcounter{table}{0}
\setcounter{equation}{0}
\setcounter{table}{0}

\renewcommand{\thesection}{\normalsize{Section \arabic{section}}}
\renewcommand{\theequation}{S\arabic{equation}}
\renewcommand{\thefigure}{S\arabic{figure}}	
	
% ===== Sec. 1 ===== %
\section{Sample preparation and structure characterization}

Polycrystalline samples of KBGB were firstly prepared by standard solid-state
reaction method as reported in~Ref.~\onlinecite{Sanders2017}. Stoichiometric
mixtures of K$_2$CO$_3$ (99.99\%), BaCO$_3$ (99.95\%), H$_3$BO$_3$
(99.99\%) and Gd$_2$O$_3$ (99.99\%) (with 6\% excess H$_3$BO$_3$ and
5\% excess of K$_2$CO$_3$ and BaCO$_3$) were thoroughly ground and
pelletized. Then the pellet was placed into an aluminum crucible and sintered
at 900$^{\circ}$C in air for 10 h. This sintering process was repeated for several
times to minimize possible impurities.

\begin{figure*}[h]
	\includegraphics[width=0.9\textwidth]{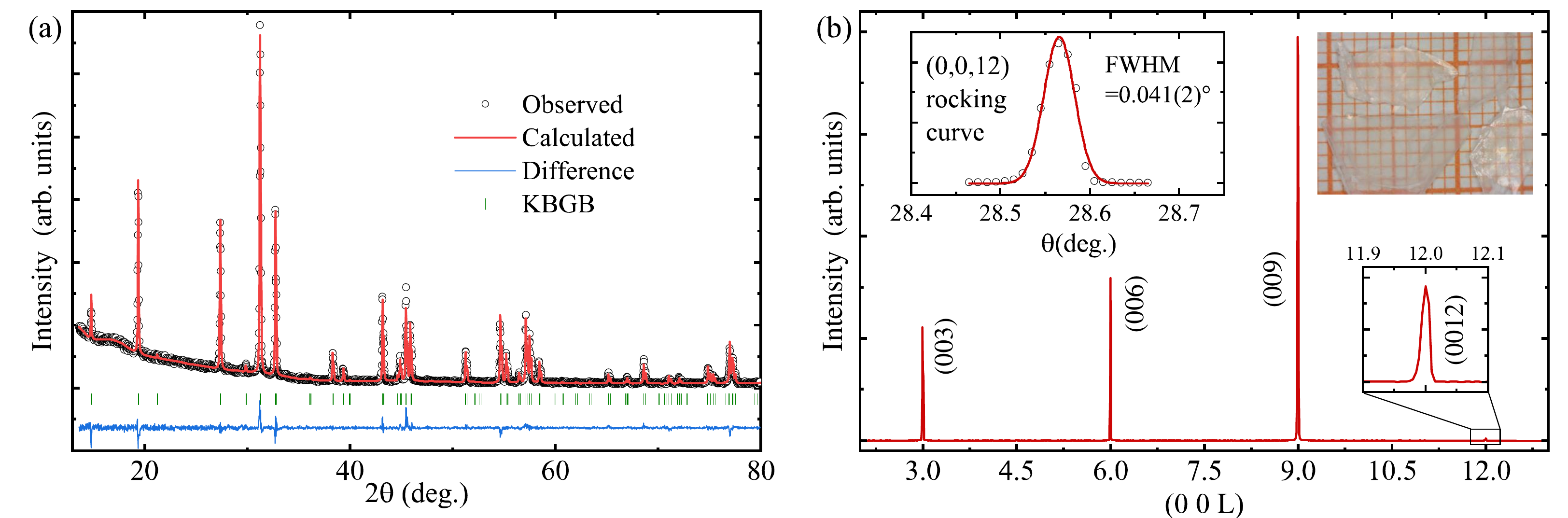}
	\caption{(a) shows the powder XRD pattern of KBGB measured at room
		temperature and corresponding Rietveld refinement. The open circle and
		red solid line represent the observed and calculated intensities,
		respectively, while the blue
		solid line shows their difference. The olive vertical bars mark the expected
		reflections for KBGB.	(b) Single-crystal XRD scan along the (0,0,L) direction
		for one representative crystal, revealing only peaks that are well indexed
		by (0,0,3n). The insets show the image of the as-grown KBGB crystals and
		the rocking-curve scan of the (0,0,12) reflection fitted by a Gaussian profile.
		The very narrow peak width of FWHM = 0.041$^{\circ}$ indicates excellent
		quality of the crystals.
	}
	\label{FigS1-XRD}
\end{figure*}

Single-crystal samples of KBGB were grown using the flux method as
reported in Ref.~\onlinecite{Guo2019KBYB}. The pre-obtained polycrystalline
KBGB with high purity was mixed with the H$_3$BO$_3$ (99.99\%)
and KF (99.9\%) fluxes in a molar ratio of 2:3:[2-3], and thoroughly ground.
The mixture was transferred into a Pt crucible, heated up to 980$^{\circ}$C
in air for 24 h, and then slowly cooled to 790$^{\circ}$C with a rate of
2$^{\circ}$C/h. After the furnace cooling, centimeter-sized crystals were
obtained on top of the fluxes.

The phase purity of the polycrystalline KBGB sample was confirmed by
powder XRD at room temperature, performed on a Bruker D8 ADVANCE
diffractometer in Bragg-Brentano geometry with Cu-K$\alpha$ radiation
($\lambda$ = 1.5406 \AA). As shown in Fig.~\ref{FigS1-XRD}(a),
the powder XRD pattern can be well fitted with the previously reported
trigonal phase of KBGB \cite{Sanders2017} ($a$ = $b$ = 5.4676(1) \AA,
$c$ = 17.9514(3) \AA), without any visible impurity peaks, indicating
high purity of the synthesized KBGB powders. The quality of the
single-crystal KBGB sample was checked by high-resolution synchrotron
XRD ($\lambda$ = 1.54564 \AA) measurements at room temperature,
performed on the 1W1A beamline at the Beijing Synchrotron Radiation
Facility (BSRF), China.
As shown in Fig.~\ref{FigS1-XRD}(b), a long $L$ scan, equivalent
to a $\theta$-$2\theta$ scan with respect to the normal direction of the 
plate-like KBGB crystal, only shows Bragg reflections well indexed by 
(0, 0, 3n) as expected for the $R$-3$m$ space group. The peak width 
(full width at half maximum, FWHM) observed in the rocking-curve scan 
of the (0, 0, 12) peak is very small, 0.041(2)$^{\circ}$, as shown in the 
inset of Fig.~\ref{FigS1-XRD}(b), which suggests excellent crystal 
quality. KBGB is relatively easy to synthesize and has excellent chemical 
stability, paving its viable way for applications in advanced cryogenics.

% ====== MCE measurements ===== %
\section{Thermodynamic and Magnetocaloric Measurements}

Comprehensive magnetothermal measurements were performed on
single-crystal samples of KBGB. The low-temperature specific heat
($C\rm_p$) and ac susceptibility ($\chi\rm_{ac}$) measurements were conducted using the Quantum Design Physical Property Measurement System (PPMS) equipped with a $^3$He–$^4$He dilution refrigerator 
(DR) insert. The specific heat data were measured under various 
out-of-plane fields ($B$//$c$) with the semi-adiabatic heat pulse method. 
The phonon contributions are negligible ($C_{\rm ph}/T \lesssim 0.003$ 
J mol$^{-1}$ K$^{-2}$ below 2~K) based on Debye $T^3$ analysis 
of high-temperature $C\rm_p$ data. The ac susceptibility $\chi\rm_{ac}$, 
as a function of temperature, was measured in zero dc field under 
different ac frequencies, with the amplitude of the ac excitation field set 
as 3~Oe. The dc magnetic susceptibility $\chi\rm_{dc}$, as a function 
of temperature down to 0.4~K, was measured using a Quantum Design 
Magnetic Property Measurement System (MPMS) equipped with 
a $^3$He insert. The isothermal dc magnetization curves in the field 
up to 7~T applied along the $a$ and $c$ axes were measured at 
0.4~K with the same setup.

\begin{figure*}[t]
	\includegraphics[width=0.95\textwidth]{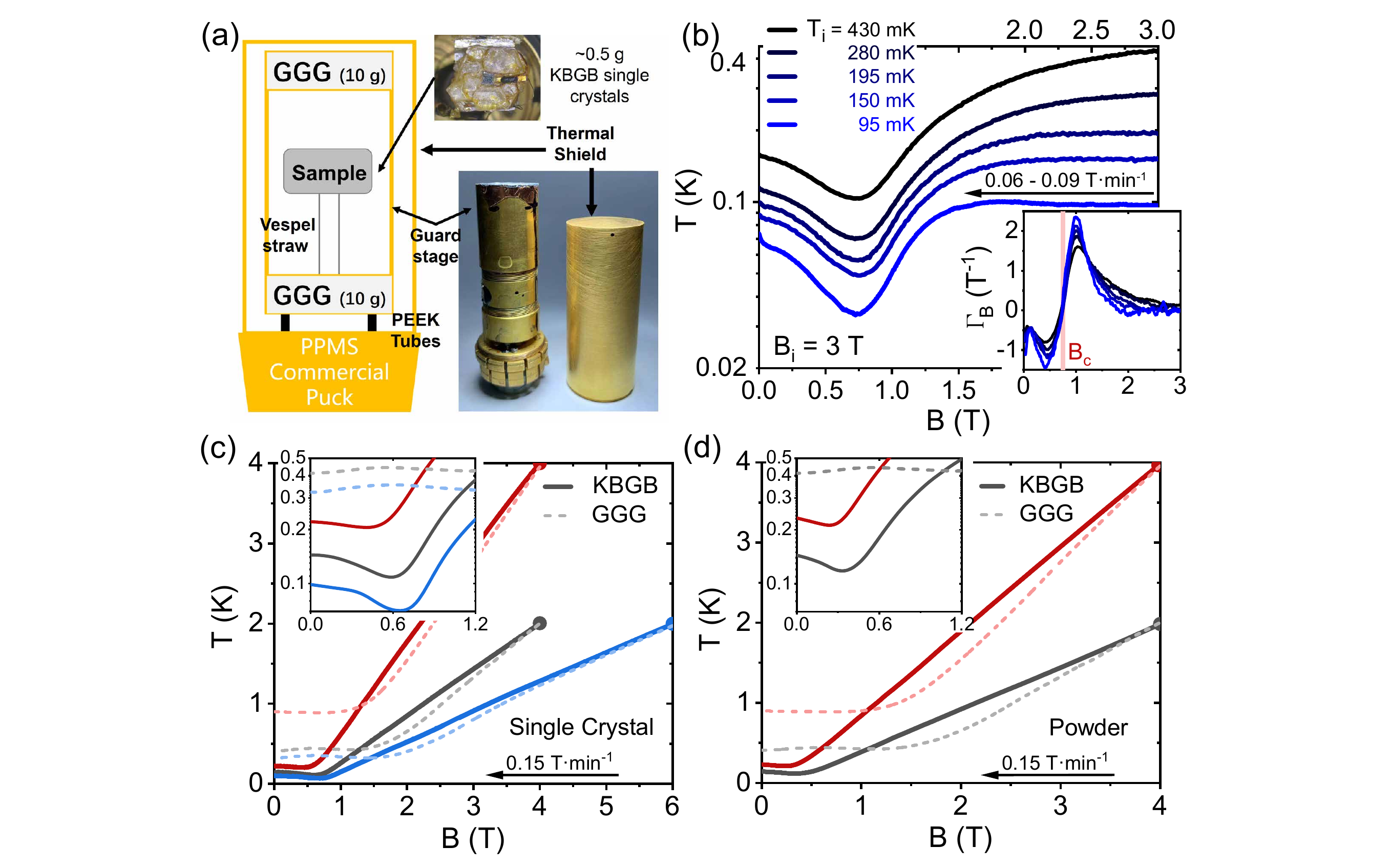}
	\caption{(a) Illustration and photo of the quasi-adiabatic two-stage
		demagnetization cooling device, KBGB single crystals used
		in the measurements are also shown. 
		(b) shows the DR-based measurements on 2.3~mg single crystal
		from an initial temperature $T_i\leq430$~mK and field $B_i=3$~T, 
		where the lowest achieved temperature is $T_m \simeq 33$~mK. 
		The inset shows the magnetic Gr\"uneisen ratio $\Gamma_B$ 
		deduced from the low-temperature isentropic $T$-$B$ lines in the 
		main plot, where the sign change is evident and the peak becomes 
		more and more pronounced as the initial temperature $T_i$ lowers. 
		(c) presents the PPMS-based measurements of isentropic curves 
		on 0.5~g KBGB single crystals starting from various initial conditions 
		($T_i=2$~K, $B_i=4$~T), (2~K, 6~T), and (4~K, 4~T), respectively, 
		where the lowest temperatures are found to be significantly lower than 
		those of GGG with the same initial conditions. The inset zooms in the 
		low-field ($B \leq 1.2$~T) and low-temperature ($T_i \leq 500$~K) regime.
		(d) Demagnetization cooling measurements of 3~g pellet consisting
		of powder samples (1.5~g KBGB and 1.5~g Ag), where the obtained
		lowest temperature are slightly higher compared to single-crystal
		KBGB sample measurements, yet much lower than that of GGG 
		[same data as in (b) panel] under the same condition. 
		The field sweep rates in the demagnetization process are 
		0.15~T/min for PPMS-based measurements ($T_i \geq 2$~K)
		and 0.06-0.09~T/min for DR-based measurements ($T_i \leq 430$~mK).
	}
	\label{FigS2:HRPuck}
\end{figure*}

Magnetocaloric effect (MCE) of the frustrated dipolar magnet 
KBGB was characterized using a homemade setup integrated 
into the PPMS, for initial temperature $2$~K $\leq T_i \leq 4$~K. 
A DR-based setup is also exploited for MCE measurements with 
low initial temperature $T_i \leq 500$~mK.

\subsection{PPMS-based setup for quasi-adiabatic demagnetization measurements}
Figure~\ref{FigS2:HRPuck}(a) shows a homemade, PPMS-based
construction for quasi-adiabatic demagnetization process, which is 
inspired by the Hagmann-Richards design of two-stage adiabatic 
demagnetization refrigeration for space applications~\cite{Hagmann1994}. 
An additional guard stage with Gd$_3$Ga$_5$O$_{12}$ (GGG) single
crystals (20~g) serve as the thermal intercept between the KBGB 
sample stage and the PPMS chamber.
In experiments, plate-like KBGB single crystals (with a total mass
of 0.5~g) are stacked along the $c$-axis and fixed on a silver foil
by cryogenic glue. We also perform demagnetization measurements
on pellet with KBGB and Ag powders 1:1 in mass, where Ag powders
are introduced to enhance the thermal conductivity. A Vespel straw 
is used to support the sample pillar inside the copper cylinder, 
which improves its thermal insulation to the PPMS chamber.
The guard stage is supported by PEEK tubes to reduce the
thermal exchange with chamber environment. The electrical
connection of the thermometer (a field-calibrated RuO$_2$ chip)
on top of the pillar is made by two pairs of twisted manganese wires
($25$ $ \mu$m in diameter and approximately $60$ cm in length)
to reduce the heat leak. A thermal shield protects the sample
from radiant heating as well as other parasitic heat loads from
the chamber. Demagnetization cooling measurements are
performed by gradually decreasing the fields from the initial field
$B_i$ at a rate of $\dot{B} = 0.15$~T$\cdot$min$^{-1}$.

The total parasitic heat load in the PPMS chamber can be estimated
from the temperature changing rate of sample after the magnet field is
exhausted, \emph{i.e.}, in the hold process with $B=0$. To be specific,
the heat load is estimated by $\dot{Q}= C_0 \dot{T}$, where $C_0$ is
heat capacity of the sample and $\dot{T}$ is the temperature change
rate. For example, when starting from an initial condition of 2~K, it is
found that $\dot{T} \approx 5 \times 10^{-6}$~K/s. Considering $C_0
\approx 0.01 $~J$/$K for 0.5~g KBGB samples in the relevant
temperature range, we thus figure out that the parasitic heat load is
$\dot{Q} \approx 0.05~\mu$W for $2$~K environment.

In Figs.~\ref{FigS2:HRPuck}(c,d) we show the isentropic lines 
of KBGB obtained through the quasi-adiabatic demagnetization 
measurements, and make a comparison with the widely used 
refrigerant GGG. The results with different initial conditions lead 
to the same conclusion, given it single-crystal or powder samples, 
that KBGB clearly outperforms GGG in the reached lowest 
cooling temperature. For example, from $T_i=2$~K and 
$B_i=4$~T, single-crystal KBGB sample (0.5~g) can reach 
as low as $T_m\simeq108$~mK, and powder sample (1.5 + 1.5~g) 
to $T_m\simeq118$~mK, both of which are much lower than
$T_m\simeq 412$~mK of GGG (20~g).

\subsection{The DR-based quasi-adiabatic demagnetization 
	measurements}

To perform MCE measurements from a lower initial temperature
below $500$~mK, a standard DR heat capacity sample mount is 
used, which provides a quasi-adiabatic condition with high vacuum 
in the $^3$He–$^4$He dilution insert of PPMS. The thermometer 
used is also RuO$_2$ semiconductor, which has been carefully 
calibrated under various magnetic fields (0-5~T) and till very low 
temperature (50~mK-4~K), further extrapolated to 30~mK according
to the scaling behavior $\ln(R-R_0) \sim  T^{-1/4}$~\cite{Tokiwa2021}.

The polymer strips are used to support the sample platform. A KBGB 
single crystal with much smaller mass of $2.3$~mg is used here, 
to avoid large magnetic torque that may break the suspended lines 
in the sample mount. To decrease the irreversible heating effect on 
the DR mount, the field sweep rate $\dot{B}$ has been reduced to 
0.06 - 0.09~T$\cdot$min$^{-1}$. Due to the small mass of the sample, 
the parasitic heat loads have a stronger influence in the DR-based 
MCE measurements (than that in the PPMS-based measurements). 
In Fig.~\ref{FigS2:HRPuck}(b), a prominent dip can be observed in 
the isentropic lines, which leads to the diverging peaks and dips with 
a sign change in the inset, again indicating clearly the existence of a 
QCP.

\begin{figure*}[htp]
	\includegraphics[width=0.9\linewidth]{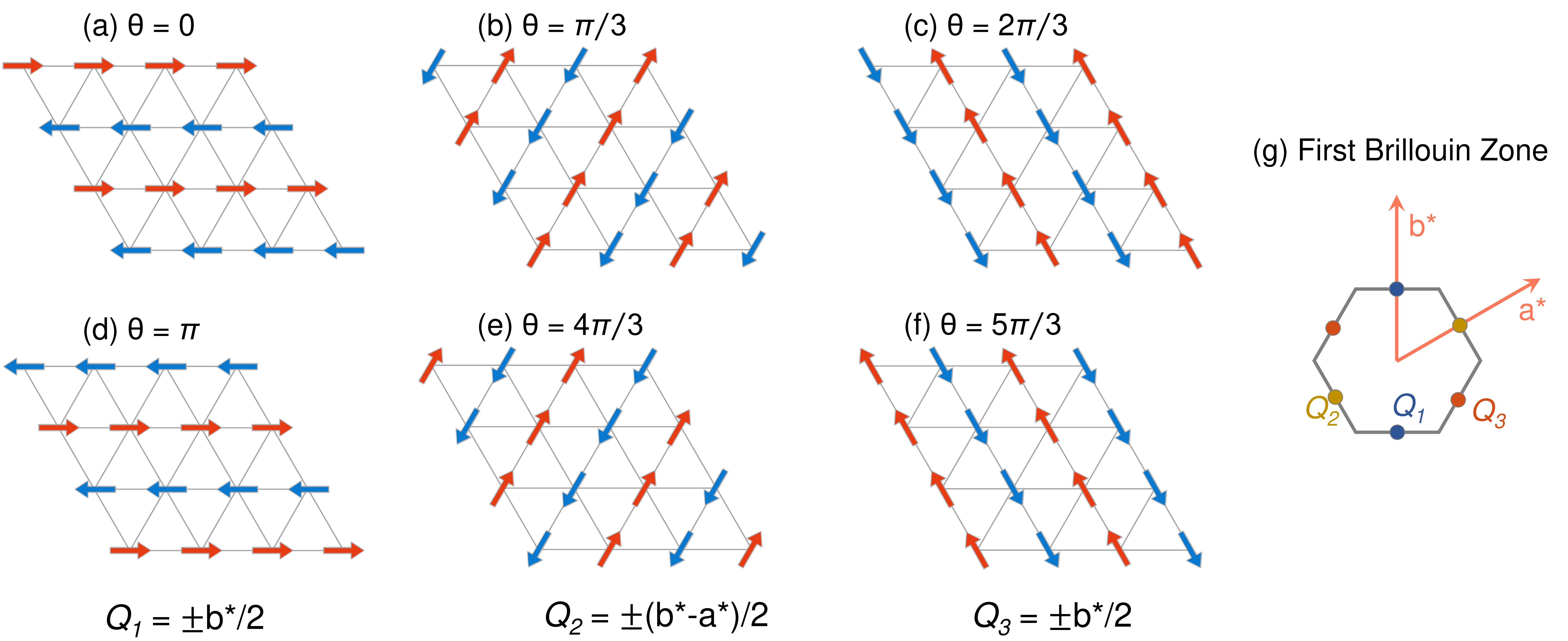}
	\caption{(a)-(f) show the magnetic configurations of the stripe order with
		6-fold degeneracy, \emph{i.e.}, 6-clock AF, which can be labeled with 
		angle $\theta$ (in complex order parameter $\Psi_{xy}$), and also by 
		ordering vector $Q_1$ (blue dots), $Q_2$ (yellow dots),  and $Q_3$ 
		(red dots) shown in (g).}
	\label{FigS3}
\end{figure*}

\section{Monte Carlo Simulations}
As the spin quantum number $S = 7/2$ is relative large, we
use the classical Monte Carlo simulations with standard Metropolis
algorithm and single spin update to study the DH model for KBGB~\cite{Sandvik2010,Overrelaxation}. The largest system size
involved in the simulations is $60\times 60$. We show the 6-fold
degenerate ground-state spin configurations in Figs.~\ref{FigS3}(a-f),
where the corresponding ordering wave vectors $Q = \pm \frac{1}{2} a^*,
\pm \frac{1}{2} b^*, \pm \frac{1}{2}(a^* - b^*)$ with $a^*, b^*$ the
primary vectors in the reciprocal lattice [see Fig.~\ref{FigS3}(g)].

In Fig.1(c-e) of the main text, we show histograms of the
complex order parameter $\Psi_{xy} \equiv m e^{i\theta}$ under magnetic
field $B=0.68$~T and at different temperatures, \emph{i.e.}, (c) $T=0.05$~K
(6-clock AF), (d) $T=0.14$~K (DSL), and (e) $T=0.25$~K (PM), respectively.
To count the histograms, we collect $5\times10^{6}$ MC samples on a
$L=12\times12$ lattice for statistics. Note the phase angle $\theta$ of the
complex $\Psi_{xy}$ can only take 6 discretized values, and the effective
theory is a 6-clock model with the order parameter $\Psi_{xy}$.

\begin{figure*}[htp]
	\includegraphics[width=0.9\linewidth]{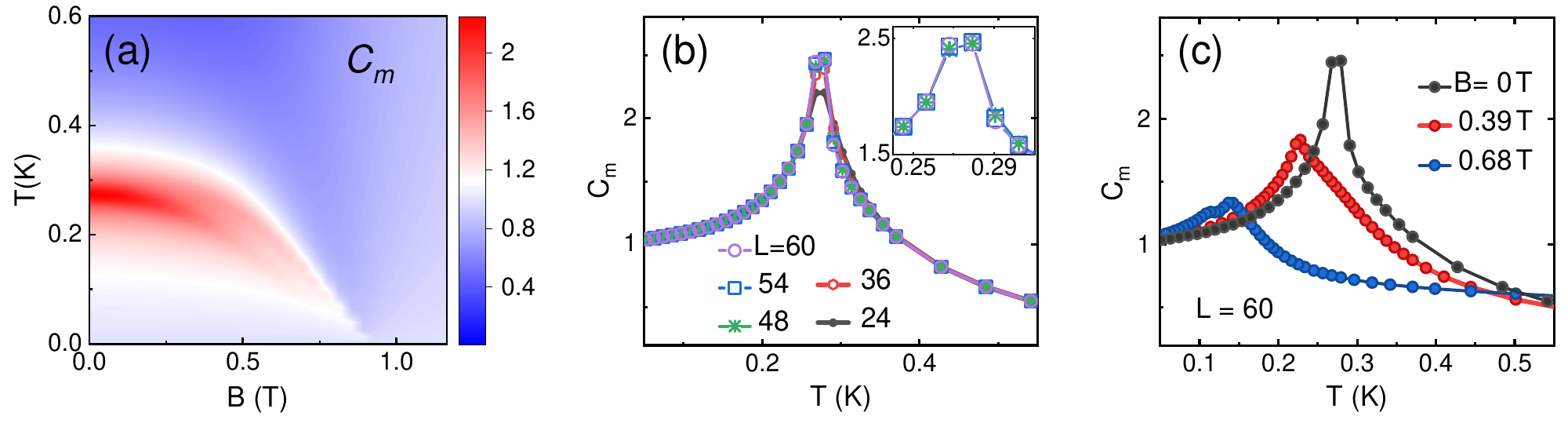}
	\caption{The calculated results of specific heat $C_m$. (a) shows the
		contour plot of $C_{m}$ data under out-of-plane fields. The $C_m$ 
		curves (b) under zero-field for different system sizes and (c) with a 
		fixed size $L=60$ and under various fields are presented. The inset in 
		(b) zooms in the $C_m$ data near the crossover temperature of about 
		$270$~mK. The MC simulations are performed on the HD model [Eq.~(1) 
		in the main text] with couplings $J=47$~mK and $D=80$~mK. }
	\label{FigS4}
\end{figure*}

The MC simulation results of specific heat are shown in Fig.~\ref{FigS4},
where the contour plot in Fig.~\ref{FigS4}(a) resembles the experimental
data in Fig.~3(b) of the main text. The peak in $C_m$ is located at about
$270$~mK, and the peak heights are converged with system sizes,
as indicated in the inset of Fig.~\ref{FigS4}(b).
In Fig.~\ref{FigS4}(c), we apply out-of-plane fields and observe
that the $C_m$ peaks move towards low-temperature side with
heights lowered, in agreement with the experimental measurements
shown in Figs.~2(a,b) of the main text.

In the MC calculations, we use the natural unit ($J=1$) and the 
following process is required for comparing the model simulations
to experiments:
(1) We replace the $\mathbf{S}_{i}$ operators in Eq.~(1)
of the main text by classical vectors, i.e.,
$\mathbf{S}_{i}\rightarrow S \mathbf{n}_{i}\equiv 7/2 \, \mathbf{n}_{i}$,
where $\mathbf{n}_{i}$ is a unit vector;
(2) The value of temperature $T$ in natural unit is multiplied by 
a factor of $J=0.047$~K;
(3) Multiply the magnetic field $B$ in natural unit (\emph{i.e.}, 
$B/JS=1$) by a factor of $Jk_{B}/(g_{c}\mu_{B})\simeq0.028$~T.

\end{document}